\documentstyle[11pt,amssymb,epsf,cite]{article}

\makeatletter
\@addtoreset{equation}{section}
\makeatother

\def\ben{\begin{equation}}
\def\een{\end{equation}}
\def\half{{\textstyle{1\over2}}}

  \let\n=\nu

\let\C=\Chi

\def\nn{\nonumber} \def\bd{\begin{document}} \def\ed{\end{document}}
\def\ds{\documentstyle} \let\fr=\frac \let\bl=\bigl \let\br=\bigr
\let\Br=\Bigr \let\Bl=\Bigl
\let\bm=\bibitem
\let\na=\nabla
\let\pa=\partial \let\ov=\overline
\newcommand{\be}{\begin{equation}}
\newcommand{\ee}{\end{equation}}
\def\ba{\begin{array}}
\def\ea{\end{array}}
\def\ft#1#2{{\textstyle{{\scriptstyle #1}\over {\scriptstyle #2}}}}
\def\fft#1#2{{#1 \over #2}}
\def\del{\partial}
\def\vp{\varphi}
\def\sst#1{{\scriptscriptstyle #1}}
\def\oneone{\rlap 1\mkern4mu{\rm l}}
\def\td{\tilde}
\def\wtd{\widetilde}
\def\ie{\rm i.e.\ }
\def\dalemb#1#2{{\vbox{\hrule height .#2pt
        \hbox{\vrule width.#2pt height#1pt \kern#1pt
                \vrule width.#2pt}
        \hrule height.#2pt}}}
\def\square{\mathord{\dalemb{6.8}{7}\hbox{\hskip1pt}}}
\newcommand{\ho}[1]{$\, ^{#1}$}
\newcommand{\hoch}[1]{$\, ^{#1}$}
\newcommand{\bea}{\begin{eqnarray}}
\newcommand{\eea}{\end{eqnarray}}
\newcommand{\ra}{\rightarrow}
\newcommand{\lra}{\longrightarrow}
\newcommand{\Lra}{\Leftrightarrow}
\newcommand{\ap}{\alpha^\prime}
\newcommand{\bp}{\tilde \beta^\prime}
\newcommand{\tr}{{\rm tr} }
\newcommand{\Tr}{{\rm Tr} }
\def\0{{\sst{(0)}}}
\def\1{{\sst{(1)}}}
\def\2{{\sst{(2)}}}
\def\3{{\sst{(3)}}}
\def\4{{\sst{(4)}}}
\def\5{{\sst{(5)}}}
\def\6{{\sst{(6)}}}
\def\7{{\sst{(7)}}}
\def\8{{\sst{(8)}}}
\def\n{{\sst{(n)}}}
\def\cA{{{\cal A}}}
\def\cF{{{\cal F}}}
\def\tV{\widetilde V}
\def\tW{\widetilde W}
\def\tH{\widetilde H}
\def\tE{\widetilde E}
\def\tF{\widetilde F}
\def\tA{\widetilde A}
\def\im{{{\rm i}}}
\def\tY{{{\wtd Y}}}
\def\ep{{\epsilon}}
\def\vep{{\varepsilon}}
\def\R{\rlap{\rm I}\mkern3mu{\rm R}}
\def\bD{{{\bar D}}}
\def\cD{{\cal D}}
\def\R{\rlap{\rm I}\mkern3mu{\rm R}}
\def\bD{{{\bar D}}}
\def\R{{{\Bbb R}}}
\def\C{{{\Bbb C}}}
\def\H{{{\Bbb H}}}
\def\CP{{{\Bbb C}{\Bbb P}}}
\def\RP{{{\Bbb R}{\Bbb P}}}
\def\Z{{{\Bbb Z}}}
\def\bA{{{\Bbb A}}}
\def\bB{{{\Bbb B}}}
\def\bC{{{\Bbb C}}}
\def\bD{{{\Bbb D}}}
\def\bZ{{{\Bbb Z}}}
\def\Re{{{\frak{Re}}}}
\def\Im{{{\frak{Im}}}}
\def\cosec{{\,\hbox{cosec}\,}}
\def\tX{{{\wtd X}}}

\def\hhtr{Hawking--Hunter--Taylor-Robinson\ }

\newcommand{\mitchell}{\it George P. \& Cynthia W.
Mitchell Institute for Fundamental Physics,\\
Texas A\&M University, College Station, TX 77843-4242, USA}
\newcommand{\tamphys}{\it Center for Theoretical Physics,
Texas A\&M University, College Station, TX 77843, USA}
\newcommand{\umich}{\it Michigan Center for Theoretical Physics,
University of Michigan\\ Ann Arbor, MI 48109, USA}
\newcommand{\upenn}{\it Department of Physics and Astronomy,
University of Pennsylvania\\ Philadelphia,  PA 19104, USA}
\newcommand{\SISSA}{\it  SISSA-ISAS and INFN, Sezione di Trieste\\
Via Beirut 2-4, I-34013, Trieste, Italy}

\newcommand{\ihp}{\it Institut Henri Poincar\'e\\
  11 rue Pierre et Marie Curie, F 75231 Paris Cedex 05}

\newcommand{\damtp}{\it DAMTP, Centre for Mathematical Sciences,
 Cambridge University\\ Wilberforce Road, Cambridge CB3 OWA, UK}
\newcommand{\itp}{\it Institute for Theoretical Physics, University of
California\\ Santa Barbara, CA 93106, USA}

\newcommand{\auth}{{\Large G.W. Gibbons\hoch{\sharp}, M.J. Perry\hoch{\sharp}
and C.N. Pope\hoch{\ddagger} } }

\begin{document}
\begin{flushright}
\hfill{DAMTP-2005-57\ \ \ MIFP-05-14}\\
\hfill{hep-th/0506233}
\end{flushright}


\begin{center}
{ \large {\Large\bf  Bulk/Boundary Thermodynamic Equivalence, and
the Bekenstein and Cosmic-Censorship Bounds for
 Rotating Charged AdS Black Holes
\\
 }}

\vspace{30pt}
\auth

\vspace{30pt}
{\hoch{\sharp}\damtp}

\vspace{3pt}
{\hoch{\ddagger}\mitchell}

\vspace{30pt}

\underline{ABSTRACT}
\end{center}

    We show that one may pass from bulk to boundary thermodynamic
quantities for rotating AdS black holes in arbitrary dimensions so
that if the bulk quantities satisfy the first law of thermodynamics
then so do the boundary CFT quantities.  This corrects recent claims
that boundary CFT quantities satisfying the first law may only be
obtained using bulk quantities measured with respect to a certain frame
rotating at infinity, and which therefore do not satisfy the first
law.  We show that the bulk black hole thermodynamic variables, or
equivalently therefore the boundary CFT variables, do not always
satisfy a Cardy-Verlinde type formula, but they do always satisfy an
AdS-Bekenstein bound.  The universal validity of the Bekenstein bound
is a consequence of the more fundamental cosmic censorship bound,
which we find to hold in all cases examined.  We also find that at
fixed entropy, the temperature of a rotating black hole is bounded
above by that of a non-rotating black hole, in four and five
dimensions, but not in six or more dimensions.  We find evidence for
universal upper bounds for the area of cosmological event horizons and
black-hole horizons in rotating black-hole spacetimes with a positive
cosmological constant.

{\vfill\leftline{}\vfill \vskip 5pt \footnoterule
{\footnotesize  \hoch{\ddagger} Research supported in part by DOE
grant DE-FG03-95ER40917 and NSF grant INTO3-24081.\vskip  -12pt}}

\pagebreak
\setcounter{page}{1}

\tableofcontents
\addtocontents{toc}{\protect\setcounter{tocdepth}{3}}
\vfill\eject

\section{Introduction} 

   There has been much progress recently in constructing solutions of
the supergravity equations describing rotating and charged black holes
in $n$-dimensional anti-de Sitter backgrounds
\cite{hawhuntay,gilupapo1,gilupapo2,d5gauge1,d5gauge2,d4gauge,d7gauge,
d5gaugeab1,d5gaugeab2}. A primary motivation for this work was the
elucidation of the thermodynamics of these black holes, with a view to
comparing it with that of the dual conformal field theories
\cite{mald,guklpo,wit} on the boundary of the spacetime, which
approaches AdS with radius of curvature $l$.\footnote{In $n=5$
dimensions, the dual boundary conformal field theory is ${\cal N}=4$
supersymmetric $SU(N)$ Yang-Mills theory, where $N^2 = \pi c^3
l^3/(2\hbar G_5)$.}  In particular the correct energies and angular
momenta for Kerr-AdS black holes, measured with respect to a frame
that is non-rotating at infinity, were calculated in all dimensions in
\cite{gibperpop}, where it was also demonstrated that these quantities
satisfy the first law of thermodynamics,
\be
dE = TdS + \Omega^i dJ_i\,.
\ee
This resolved some of the apparent ambiguities in earlier work, that had
focused on energies and angular velocities measured with respect to a
particular frame rotating at infinity, which we shall denote with
primes throughout this paper.\footnote{Not all quantities measured in
the rotating coordinate system are changed from their values in the
asymptotically static frame, but for clarity we denote all quantities
measured in the rotating frame with primes in the present discussion.}
As shown in \cite{gibperpop}, these do not satisfy the first law of
thermodynamics:
\be
dE' \ne  T'dS' + {\Omega^i}' dJ_i'\,,
\ee
since the asymptotic rotation rate in this frame depends on the black-hole
rotation parameters.

    In a recent paper, Cai et al. \cite{cai} noticed that by passing from 
the bulk quantities $(E', J_i')$ to
the dual CFT quantities $(e', j_i', \ldots)$ on the boundary, and 
by including an additional pressure term $p'$ and suitably-defined volume
term $v'$, the equation
\be 
de' = t' ds' + {\omega^i}' dj_i' - p' d v'
\ee
holds.  They interpreted this equation as the first law for the dual
CFT, and furthermore they made the surprising claim that no such
analogous CFT thermodynamic variables can be introduced that are dual to the
unprimed bulk quantities, and which satisfy the first law.
\be
de = t ds + \omega^i dj_i - p d v  \,.\label{cftlaw2}
\ee
One purpose of this paper is to refute this surprising claim, and on
the contrary to demonstrate that following the perfectly standard
transcription rules relating bulk and boundary quantities, the first
law (\ref{cftlaw2}) does indeed hold.  We also raise questions as to
whether the interpretation of the primed quantities given in
\cite{cai} is physically correct.

   A remarkable feature of \cite{cai}, following earlier work in
\cite{klemm1,klemm2}, is the observation that the primed bulk quantities
satisfy a Cardy-Verlinde \cite{verlinde} type formula, 
\bea
S' &=& \fft{2\pi l}{n-2}\, \sqrt{E_c' (2E' - E_c')}\,,\nn\\
E_c' &=& (n-1) E' - (n-2)( T' S' + {\Omega^i}' J_i')\,,
\eea
whereas the unprimed quantities measured with respect to a
non-rotating frame at infinity do not.  This has motivated us to
re-examine the old question of whether or not AdS black holes satisfy
some sort of possibly modified Cardy-Verlinde formula, and a
Bekenstein-type bound of the form\footnote{Note that the Bekenstein
bound does not contain Newton's constant, and so it makes sense for
any thermodynamic quantity in an AdS background.  However, in this
paper our concern is solely with black holes.}
\be
E \ge \fft{(n-2) S}{2\pi l}\,.
\ee

   The existence of the Bekenstein bound is a necessary but not
sufficient condition for the validity of a Cardy-Verlinde formula.  In fact, 
we find that while there appears to be no universal formulation of a modified
Cardy-Verlinde formula that will cover all of the charged and rotating 
AdS black holes that we know of, we do find that a Bekenstein bound holds 
in all cases.  

    In fact, the Bekenstein bound follows from a stronger and more
fundamental inequality, the cosmic censorship bound, which takes the
form
\be
E \ge \fft{(n-2) A}{16\pi l}\, \Big[
     l\, \Big(\fft{A}{ {\cal A}_{n-2} }\Big)^{-\ft1{n-2}} +
\fft1{l}\,  \Big(\fft{A}{{\cal A}_{n-2} }\Big)^{\ft1{n-2}} \Big]\,,
\label{ccbound0}
\ee
where $A$ is the area of the event horizon, or more generally, in 
time-dependent cases, the area of the outermost apparent horizon, 
and ${\cal A}_{n-2}$ is the volume of the unit $(n-2)$-sphere.
We show that the recently-constructed exact solutions for rotating
and charged AdS black holes give strong support for the conjectured
cosmic censorship bound.

   As well as lower bounds for the energy in terms of the entropy, it
is well known that there are interesting upper bounds for the
temperature as a function of entropy, for black holes in
asymptotically flat spacetimes.  It turns out that these may be generalised 
to the static anti-de Sitter case, taking the form
\be
4\pi T\le  (n-3) \Big(\fft{A}{{\cal A}_{n-2} }\Big)^{-\ft1{n-2}} +
                 (n-1) l^{-2} \Big(\fft{A}{{\cal A}_{n-2} }\Big)^{\ft1{n-2}}\,,
\ee
in $n$ dimensions.  That is, the temperature is never greater than the
value it would have in the Schwarzschild-AdS solution with the same
entropy.  We investigate whether the bound extends to stationary black
holes, finding that it is obeyed by all Kerr-AdS black holes in four
and five dimensions, but not in dimensions six or higher.

     The uncharged rotating black hole solutions are
of course valid also if the cosmological constant is taken to be
positive, corresponding to sending $l^2\rightarrow - l^2$.  In this
case, an additional, cosmological, horizon is present.  We verify that
these solutions support the general conjecture that the area $A_C$ of
the cosmological horizon satisfies
\be
A_C \le {\cal A}_{n-2}\, l^2\,,
\ee
and the black hole horizon satisfies the inequality
\be
A_H \le {\cal A}_{n-2}\, l^{n-2}\, \Big( \fft{n-3}{n-1}\Big)^{\ft{n-2}{2}}\,.
\ee

   The plan of the paper is as follows.  In section 2 we establish a
general equivalence between bulk and boundary thermodynamics, and
explain our disagreement with some of the work in \cite{cai}.  In
section 3, we review the Cardy-Verlinde formula and its consequence,
the AdS-Bekenstein bound, explaining why it is saturated at the
Hawking-Page phase transition.  We show that a simple modification
holds for Reissner-Nordstr\"om-AdS black holes, and gives rise to a
strengthened form of the Bekenstein bound.  We also show that in all
dimensions rotating black holes without charge satisfy the
AdS-Bekenstein bound.  In section 4, we examine a large number of
examples of rotating and/or charged black holes, finding that despite
the failure in general of the Cardy-Verlinde formula, the
AdS-Bekenstein bound, or its strengthened electrostatic form, holds.
In section 5 we discuss how the AdS-Bekenstein bound may be regarded
as a consequence of the AdS cosmic censorship bound, and demonstrate
in all the examples we have checked that the AdS cosmic censorship
bound does indeed hold, in some cases strengthened by an electrostatic
contribution.  Section 6 discusses upper bounds for the temperature of
AdS black holes, in terms of their entropy.  We find that rotating
black holes in four and five dimensions always have a temperature that
is less than that of the Schwarzschild-AdS solution with the same
area.  However, rotating black holes of dimension six or higher do not
satisfy such a bound.  Section 7 contains a brief discussion of the
areas of the cosmological and black-hole horizons for rotating black
holes with positive cosmological constant.  We collect for the
reader's convenience, in an appendix, the pertinent formulae for
Kerr-AdS black holes in arbitrary dimensions.  Our conclusions are
contained in section 8.

\section{Bulk and Boundary Thermodynamics}

   The purpose of this section is to show that there is a universal
rule allowing one to pass between bulk and boundary quantities in 
such a way that if one set of quantities satisfies the first law
of thermodynamics, then so will the other.  

\subsection{Tolman or UV/IR scaling transformations}\label{yscal}

   It is quite generally true
 that if an arbitrary  thermodynamic  system satisfies
the first law of thermodynamics without a pressure term,
\ben
dE= TdS +\Omega^i dJ_i + \Phi_i dQ_i\,,
\een
then associated with it is a second system,
with pressure equal to the energy density divided by the spatial dimension,
which satisfies the first law with pressure term:
\ben
de=tds+\omega^i dj_i + \phi_i dq_i -pdv\,.
\een
Actually, since the first system need have no natural
dimension associated to it, the spatial dimension $(n-2)$ of the second
system can be arbitrary.  The thermodynamic quantities of the
second system, denoted by lower-case letters, are related to those of 
the first system by
\ben 
\fbox {$\begin{array}{rcl}
e &=& {l \over y} E\,, \qquad
\omega^i = \fft{l}{y} \Omega^i\,,\qquad \phi_i=\fft{l}{y} \Phi_i\,,\qquad
       t = \fft{l}{y} T\,,\\
 s &=& S\,,\qquad\quad  j_i = J_i\,,\qquad \quad q_i= Q_i\,, 
\label{good}
\end{array}$} 
\een
with
\be
v = {\cal A}_{n-2}\, y^{n-2}\,,\qquad p = \fft{e}{(n-2)\, v}\,,
\label{vpdef}
\ee
where ${\cal A}_{n-2}$ is the volume of the unit $(n-2)$-sphere.  Here,
$y$ is to be thought of as the radius of the second system, and $l$
is an arbitrary constant, which in our application is related to the 
cosmological constant, so that $R_{\mu\nu}= -(n-1) l^{-2}\, g_{\mu\nu}$.
Note that in (\ref{good}) the intensive quantities $(T,\Omega^i,\Phi_i)$
are scaled, as is the energy $E$, whilst the extensive quantities 
$(S,J_i,Q_i)$ are not scaled.

    As it stands, the above result is a mathematical triviality.
However, in the case we are considering, where the first system is a
rotating charged black hole in an AdS$_n$ background, it allows us to
relate the bulk thermodynamic quantities associated with the black
hole to the boundary quantities associated with the dual conformal
field theory.  The quantities $(E,T,S,\Omega^i,J_i,\Phi_i,Q_i)$ are
all evaluated with respect to a coordinate frame $(t, y, \hat\mu_i,
\varphi_i)$ that is non-rotating and asymptotically spherical at
infinity.\footnote{The time coordinate $t$ should not be confused with
the CFT temperature $t$ --- it should be clear from the context which is
which.}  In these coordinates, the metric of the rotating black hole
approaches the AdS metric
\be
d\bar s^2 = - (1+ y^2 l^{-2}) dt^2 + \fft{dt^2}{1+ y^2 l^{-2}}
  + y^2 \sum_{k=1}^{N+\ep} (d\hat\mu_k^2 + \hat\mu_k^2 d\varphi_k^2)\,,
\ee
in $n=2N+\ep+1$ dimensions, where $\ep=(n-1)$ mod 2 (see
\cite{gibperpop}, and appendix A, for a more detailed discussion).
Note that the radial coordinate $y$ is related to the Fefferman-Graham
coordinate $z\sim l^2/y$ for which the metric asymptotes to $-l^2
dt^2/z^2 + l^2 dz^2/z^2 + l^4/z^2 d\Omega_{n-2}^2$.

    The 
Killing vector $\del/\del t$ is normalised so that near infinity
\be
g(\fft{\del}{\del t},\fft{\del}{\del t}) \rightarrow 
    -\fft{y^2}{l^2}\,,
\ee
and this fixes the normalisation of the quantities $(E,T,\Omega^i,\Phi_i)$.
A boundary conformal field theory living on a surface $y=$constant will
thus occupy the volume $v$ given in (\ref{vpdef}).  The intensive 
quantities $(t,\omega^i,\phi_i)$ of the CFT are then given by the 
standard Tolman redshifting factor, or, in the language of the AdS/CFT
correspondence, the UV/IR connection, which coincides with our 
formulae in (\ref{good}).  The pressure $p$ is that expected of a 
conformally-invariant system, the trace of whose energy-momentum tensor
should vanish.  One reason why the extensive quantities $S$, $J_i$ and
$Q_i$ cannot scale under the UV/IR connection is that they are subject
to quantisation conditions, and are given by integers.  

   The upshot of the above discussion is that the introduction of the
pressure term is a triviality, which ensures that if the first law
of thermodynamics holds in the bulk, then it holds also in the boundary
CFT.

\subsection{Relation to earlier work}

   Although for us the introduction of the pressure term is, as we
have explained above, a triviality, because our bulk quantities
satisfy the first law of thermodynamics, it is less transparent if
bulk thermodynamic variables are chosen that do not satisfy the first
law.  As we showed in \cite{gibperpop}, the way to obtain bulk
thermodynamic quantities for black holes that satisfy the first law is
by calculating them with respect to a frame that it non-rotating at
infinity.  The energy measured in this frame can be derived
\cite{gibperpop} using the Ashtekar-Magnon-Das conformal definition of
mass in AdS backgrounds \cite{ashmag,ashdas}.  It has also been shown
\cite{derkat} that the same expression can be derived from the
superpotential of Katz, Bi\v c\' ak and Lynden-Bell \cite{kabily}.  
A further calculation leading to the same expression for the energy
was given recently in \cite{deskantek}.

     There are, of 
course, infinitely many frames one could choose that do rotate, with
different rotation rates, at infinity.  One popular choice is the 
asymptotically-rotating coordinate system in which Carter first 
wrote the Kerr-AdS black hole in four dimensions \cite{carter}. Analogous 
rotating frames were introduced in five dimensions by 
Hawking, Hunter and Taylor-Robinson \cite{hawhuntay}, and in all higher 
dimensions in \cite{gilupapo1}.  In these papers, the metrics
are given in a coordinate system which is rotating with angular 
velocity  
\be
\Omega_\infty^i = \fft{a_i}{l^2} \label{ominf}
\ee
with respect to an asymptotically static frame,
where $a_i$ are the rotation parameters. (See appendix A for a 
summary of the salient details of the Kerr-AdS metrics.  In the appendix,
the metrics are given in an asymptotically static coordinate system.)  

    The geometrical significance of this particular rotating frame is that with
respect to it the Kerr-Schild congruence, which was used to construct
the solution, is non-rotating at infinity.  However, this in itself
does not appear to endow it with any privileged dynamical
significance.  Nevertheless one can certainly, as has been done in
some of the literature, associate with it energies and angular
velocities, which we shall denote by primes, that are given in terms of
the unprimed non-rotating thermodynamic quantities by \cite{gibperpop}
\be
E'= E -\fft{a_i}{l^2}\, J_i\,,\qquad {\Omega^i}' = \Omega^i - 
         \fft{a_i}{l^2} \,,
\ee
with all the other quantities being the same in the primed and the
unprimed frame.\footnote{The reason why $J_i=J_i'$ is that ``passing
to the rotating frame'' means in effect choosing a new timelike
Killing field from which $E'$ is constructed, but retaining the same
angular Killing fields from which the $J_i$ are constructed.  In other
words, one introduces the new rotating coordinates $(t',\varphi_i')$,
related to the asymptotically static coordinates $(t,\varphi_i)$ by
$t'=t$, $\varphi_i'=\varphi_i - a_i l^{-2} t$, and associates the energy
$E'$ with the Killing vector $\del/\del t'=\del/\del t + a_i l^2
\del/\del\varphi_i$, as opposed to the energy $E$ associated with
$\del/\del t$.  Thus passing to a rotating frame is {\sl not} the same
as performing an asymptotic $SO(n-1,2)$ transformation; it is merely
picking a new basis for the Lie algebra $\frak{so}(n-1,2)$.}  Note
that
\be
E'- {\Omega^i}' J_i' = E - \Omega^i J_i = E' - {\Omega^i}' J_i\,.
\ee
Throughout the rest of this paper, we shall use the symbols $E'$ and
${\Omega^i}'$ to denote energies and angular velocities measured with
respect to the asymptotically rotating frames for which (\ref{ominf}) holds.

    Although $E'$ appears to have no special physical significance, it
turns out that it provides a useful bound for the true energy $E$, in 
other words
\be
E \ge E'\,,
\ee
with equality if and only if the black hole is non-rotating.

   As we noted in \cite{gibperpop},
\be
dE' \ne T dS + {\Omega^i}' dJ_i\,.
\ee
However, Klemm et al. \cite{klemm1,klemm2} discussed 
the thermodynamics of rotating AdS black holes with a single non-vanishing
rotation parameter, and obtained an extended system involving a chemical
potential $\mu$ and number $N$, satisfying the first law.  More 
recently, Cai et al. \cite{cai} have introduced thermodynamic quantities 
which in 
our notation we shall write as $e'$, $t'$, $s'$, ${\omega^i}'$, $j_i'$,
$p'$, $v'$, given by
\be
v'= \fft{{\cal A}_{n-2}\, r^{n-2}}{\prod_j \Xi_j}\,,\qquad
 p' = \fft{e'}{(n-2)\, v'}\,,\label{ppvv}
\ee
\ben
\begin{array}{rcl}
e' &=& {l \over r} E'\,, \qquad
{\omega^i}' = \fft{l}{r}\, {\Omega^i}'\,,\qquad
       t' = \fft{l}{r} T\,,\qquad
s' = S\,,\qquad\quad \,  j_i' = J_i\,.
\label{bad}
\end{array}
\een
and they have shown that these satisfy the first law 
\be
de'= t' ds' + {\omega^i}' dj_i' - p' dv'\,.\label{primefirst}
\ee

   Note that the formula (\ref{ppvv}) for $v'$ is not as we defined in
(\ref{vpdef}), but rather has the additional factor $\prod_i \Xi_i$ in the
denominator.  This is needed in order to get the first law
(\ref{primefirst}) in the primed variables to work out, to compensate
for the failure of the bulk primed quantities to satisfy the first
law.  It is also not difficult to see that if one chooses a different
rotation rate at infinity, replacing the right-hand side of
(\ref{ominf}) with some general functions of the rotations $a_i$, then
one cannot in general find a formula for $v'$ of the form (\ref{ppvv})
with the factor $\prod_i \Xi_i$ replaced by a suitable function of the
$a_i$.  To that extent, the fact that the primed CFT quantities
satisfy the first law (\ref{primefirst}) is no entirely fortuitous.
Nevertheless, it seems to us that it is the unprimed CFT quantities
given by (\ref{good}) and (\ref{vpdef}) that most closely correspond
to the physical situation that motivated the work in \cite{hawhuntay}.
In other words, the relevant CFT should rotate, relative to a frame
non-rotating at infinity, with the {\it same} angular velocity as that
of the black hole in the bulk theory.  

    One could pass to a frame that
is co-rotating with the black hole, i.e. one whose angular velocity 
with respect to the frame that is non-rotating at infinity is equal to
$\Omega^i$, given by (\ref{omapp}).  The associated Killing vector 
\be
K=\fft{\del}{\del t} + \Omega^i \fft{\del}{\del\varphi^i}
\ee
(expressed using the asymptotically non-rotating coordinates in (\ref{bl}))
coincides on the horizon with its null generator, and is, provided
$|a_i| <l$, everywhere timelike outside the horizon.  This has the
desirable feature that local energy densities measured with respect to
this Killing vector are everywhere positive \cite{hawkreal}.  However,
it has the distinct disadvantage that when considering any energy
exchange between the rotating black hole and its environment, one must
change to a new rotating frame because in general $\Omega^i$ changes.
It is for this reason that the first law of thermodynamics does not
hold with respect to the primed quantities.  More generally, one could
consider a Killing vector of the form
\be
\wtd K= \fft{\del}{\del t} + \wtd\Omega^i \fft{\del}{\del\varphi^i}\,.
\ee
A simple calculation shows that on the horizon,
\be
g(\wtd K, \wtd K) = 
g_{ij} (\wtd \Omega^i - \Omega^i)(\wtd \Omega^j-\Omega^j)\,,
\ee
where $g_{ij}= g(\del/\del\varphi_i,\del/\del\varphi_j)$, and thus we
see that for any angular velocity $\wtd\Omega^i$ that differs from
$\Omega^i$, the associated Killing vector $\wtd K$ is spacelike on
(and therefore in the neighbourhood of) the horizon.  In particular,
this applies to the choice $\wtd\Omega^i=\Omega_\infty^i =
a_i/l^2$.  Thus to use the primed frame would neither achieve
positivity of the local energy density nor a simple form for the first
law.  It seems, therefore, that neither it, nor any other frame that
is rotating at infinity (other than, possibly, the frame that is
rotating with the angular velocity of the black hole) has any
particular merit or advantage over the frame that is non-rotating at
infinity.  Of course physical results cannot depend upon an arbitrary
choice of frame.\footnote{It should be emphasised that the different 
expressions for the energy and angular velocity in different frames is
not the result of the coordinate transformation {\it per se}, but of using the 
transformed  time coordinate when defining the  energy and angular velocity.}
  It is clear that we can describe a rigidly-rotating
gas either as being at rest in a rotating frame, or moving in a
non-rotating frame.  The choice which seems to us the most
straightforward and simple is the latter.  Similarly, the calculations
in \cite{nzn} need not have been done using the primed quantities, and
we disagree with the assertion in \cite{nzn} that it is necessary to
use the primed quantities in order ``to extract data useful to a dual
CFT.''  As we have seen above, this is a trivial matter using the
non-rotating frame.

   Using the primed energy $E'$ is precisely analogous to using the
kinetic energy of a particle with respect to a rotating frame, such as
that of the earth.  It can be done, but it is then necessary to
consider the contributions to the energy and the equations of motion
due to the centrifugal and Coriolis forces.  If the particle is freely
moving on the rotating platform, and the rotation rate is changed, the
kinetic energy with respect to the rotating frame will obviously
change, while it will clearly be constant with respect to an inertial
frame.  There would seem to be no merit in introducing an artificially
time-dependent rotating frame merely to describe straight-line motion
in inertial coordinates.  If instead of free particles we considered a
gas in a state of rigid rotation, we would have a situation rather
more analogous to that of the CFT.  The gas would exert a pressure on
its container, which in principle could be measured in any rotating
frame, but the two that are most relevant are surely the rigid
rotating frame co-rotating the gas, or the one that is non-rotating
with respect to an inertial coordinate system.  As we have explained
earlier, if the rotation rate changes with time, it is the latter
which is more convenient.  Choosing to use the energy $E'$ in the
rotating black-hole problem is the equivalent of using a frame that is
neither non-rotating at infinity nor is it rotating at the angular 
velocity of the black hole horizon.  Furthermore, in previous work where
the energy $E'$ has nevertheless been used, the necessary corrections
to compensate for the changes in the rotation rate of the primed frame have
been omitted.

   In \cite{cai}, the CFT is assumed to be on a surface of large $r$
in Boyer-Lindquist coordinates, and the spatial volume is supposed to
be the volume of that surface.  It should be noted that although this
spatial surface has the topology of an $(n-2)$-sphere it does not have
an $SO(n-1)$ isometry group even asymptotically at large $r$.  If
one nevertheless chooses this $r=$constant boundary one must face up to
the fact that the temperature will be space dependent and there will
be no conventional thermodynamic interpretation where a global temperature 
is well defined.

   For example, in four dimensions the Kerr-AdS metric at large $r$
approaches the form
\be
ds_4^2 = \fft{r^2 \Delta_\theta}{l^2\, \Xi} \, \Big[
 - dt^2 + l^2 \Big\{\fft{\Xi\, d\theta^2}{\Delta_\theta^2}  +
  \fft{\sin^2\theta}{\Delta_\theta} (d\phi + a l^{-2}\, dt)^2\Big\}\Big]\,,
\ee
where
\be
\Delta_\theta = 1 - \fft{a^2}{l^2} \cos^2\theta\,,\qquad
\Xi=1-\fft{a^2}{l^2}\,.
\ee
Defining a new coordinate $\hat\theta$ by $\tan\hat\theta = (\tan\theta)/
\sqrt{\Xi}$, it can be seen that the 2-metric enclosed in braces is
nothing but the standard unit 2-sphere, whose volume is of course 
$4\pi$.  The metric in the square brackets is that of a three-dimensional
rotating Einstein universe.  The CFT metric is in fact conformal to this,
and with respect to this metric the spatial volume is
\be
\fft{2 \pi r^2}{\Xi}\int_0^\pi d\hat\theta \sin\hat\theta \Delta_\theta
= \fft{4\pi r^2 l}{a\sqrt{\Xi}} {\rm{arcsin}}(a/l)\,.
\ee
This is not equal to $4\pi r^2/\Xi$, which is the value given in
\cite{cai}.  We are thus unsure as to precisely which spatial volume 
the quantity $4\pi r^2/\Xi$ in \cite{cai} is supposed to be.  Similar
remarks apply to all the higher-dimensional expressions for $v'$ given in
\cite{cai} and reproduced in (\ref{ppvv}).

   A striking feature of the work in \cite{cai} is the finding that
the quantities $e'$, $t'$, $s'$, ${\omega^i}'$, $j_i'$, $p'$ and $v'$
satisfy a suggested formula of E. Verlinde \cite{verlinde}, which
itself was based on an attempt to incorporate Bekenstein's conjecture
\cite{bekenstein} of some sort of bound relating entropy, energy and
radius.  This has motivated us to look in more detail at the general
question of such formulae and bounds, which we shall do in the next
section.

\section{The Cardy-Verlinde Formula and the Bekenstein Bound}

\subsection{The ideal Cardy-Verlinde gas}

     According to a proposal of E. Verlinde, a CFT living
on an $(n-1)$-dimensional Einstein Static Universe (ESU) of radius $y$
and hence metric
\ben
ds^2 =-dt^2 +y^2 d \Omega _{n-2}^2\,,
\een
where $d \Omega _{n-2}^2$ is the canonical round metric on
$S^{n-2}$ should have

\begin{itemize}

\item A pressure, energy and volume related by
\ben
p= { 1 \over n-2} {e \over v}\,,\qquad v={\cal A}_{n-2} y^{n-2} \,,
\een 

\item
 An entropy $s$ given by
\ben
s= {2 \pi y \over n-2} \sqrt{e_c(2e-e_c)}\,,\label{cvform}
\een
where $e_c$ is a measure of the extent to which the energy $e$
 is non-extensive
and given by
\ben
e_c := (n-2) (  e-ts-\omega j -\phi q +pv )\,. \label{CardyVerlinde}
\een

\end{itemize}

   Subsequent work \cite{kutlar} showed that for free field theory,
the Cardy-Verlinde formula does not hold exactly, but it agrees, up to
a constant factor, in the high-temperature limit.  That is, at large
$T$ one has $E_c \propto T^{n-3}$, and $E \propto T^{n-1}$, and hence $S
\propto T^{n-2}$. However the factor of proportionality is wrong.
Nevertheless, it is still possible that it, or some modified form, may
hold in the strongly-interacting case that is relevant for the AdS/CFT
correspondence.

   In this limit, it is more convenient to discuss bulk, rather than
boundary, quantities.  As we have emphasised in section \ref{yscal},
one can freely translate back and forth between the two descriptions.
Using the Tolman redshifting formula, or the UV/IR relation between
lower case and capital letter quantities, one sees that (\ref{cvform})
may be re-written in terms of bulk quantities as
\ben
S= {2 \pi l \over n-2} \sqrt{E_c(2E-E_c)}\,, \label{CardyVerlinde2}
\een
where $E_c$ is given by
\ben \fbox{$\displaystyle
E_c := (n-2) \Bigl
[  (1+ {1 \over n-2}   )
 E-TS-\Omega J -\Phi Q \Bigr ]\,. \label{CardyVerlinde1} $}
\een
and we identify $E$ as the conformal generator
of  $ J_{0n} \in \frak{so}(n-1,2)$  associated to
the asymptotically-static Killing field  $ \partial \over \partial t$,
\ben
E= l^{n-3}  J_{0n}\,.
\een
For later purposes, we  re-write (\ref{CardyVerlinde1}) as\footnote{The 
symbol $:=$ means that the quantity
on the left-hand side is defined by the expression on the right-hand side.
In particular we shall, for the sake of  clarity, {\sl always} stick with this
primary definition
of $e_c$, and the analogous blue shifted quantities $E_c$. If we need
to modify the definition we will indicate any modified quantity
by a circumflex, thus $\hat e_c$. The importance of this
cannot be over-emphasised. If one does not stick to the
primary definition (\ref{CardyVerlinde1a}), then the
question of the existence or non-existence of such a formula becomes
completely meaningless, since one could always define $e_c$ in such
a way that the Cardy-Verlinde formula became a trivial identity!}
\ben
E_c= (n-1)   E- (n-2) \bigl [ TS+\Omega J -\Phi Q\bigr ] \,.
\label{CardyVerlinde1a}
\een

   From now on we shall primarily be concerned with the
Cardy-Verlinde formula in terms of ``bulk,'' that is black hole,
quantities. However, it is worth remarking that in some papers the
anti-de Sitter radius $l$ is replaced by $r_+$, where typically $r_+$
is the radius of the horizon in Schwarzschild or, in the rotating
case, Boyer-Lindquist coordinates. In effect this amounts to setting
$y=r_+$ in the redshifted CFT form. However, unless $r_+>> l$, this
cannot be regarded as a legitimate application of the UV/IR relation,
since if $r_+ \sim l$ then $-g_{tt}$ is not well
approximated by $y/l$.

\subsection{Anti-de Sitter Bekenstein bound and Hawking-Page transition}

    The (strict) Cardy-Verlinde formula (\ref{CardyVerlinde2})
may be regarded as a formula
for the energy $E$ in terms of the entropy $S$ and
non-extensive energy $E_c$:
\ben
E= {1 \over 2} E_c +
{ 1 \over 2 E_c} \Big[ { (n-2) S \over 2 \pi l } \Big]^2
\,.\label{CardyVerlinde3} 
\een
Minimization with respect to $E_c$ leads to the lower bound
\ben \fbox{$\displaystyle
2 \pi l E\ge (n-2) S\,.\label{Bekenstein}$}
\een
which we shall refer to as the {\it Anti-de Sitter Bekenstein Bound},
regardless of whether it arises from a Cardy-Verlinde formula.
Clearly the anti-de Sitter Bekenstein bound is a necessary,
{\it but not sufficient} condition for the existence of
a Cardy-Verlinde formula.

   This lower bound for the energy in terms of the entropy,
or alternatively upper bound for the entropy in terms of the
energy, is attained when
\ben
E=E_c= {(n-2) S \over 2 \pi l}\,.
\een
Note that for $E > {(n-2) S \over 2 \pi l}$ there are two values
of $E_c$ satisfying the Cardy-Verlinde formula (\ref{CardyVerlinde3}),
whilst if $E < {(n-2) S \over 2 \pi l}$, there are none.

    The calculation above may be re-organised as follows.
The thermodynamic potential $\Psi$ of the bulk black hole
is given by
\ben
\Psi=E-TS-\Omega J -\Phi Q\,.
\een
Thus
\ben
E_c= E+  (n-2) \Psi\,,\qquad 2E-E_c =E-(n-2)\Psi\,.
\een
The Cardy-Verlinde formula can be cast in the Pythagorean form
\ben
{E^2 \over (n-2) ^2 }  =
{S^2 \over  4 \pi ^2l^2 } + \Psi^2\,. \label{Pythagoras}
\een

    We see from (\ref{Pythagoras}) that the Bekenstein
bound is attained if and only if the thermodynamic potential
vanishes,
\ben
\Psi=0\,.
\een
If one accepts the quantum statistical relation
between thermodynamic potential and Euclidean action $I$,
\ben
\Psi =T I\,,
\een
then the Bekenstein bound is attained when the Euclidean action
vanishes.  In the black hole case, this indicates that the Euclidean
black hole solution with large energy $E$ no longer has smaller action
than that of flat space, and a type of phase transition is indicated,
as was first discussed by Hawking and Page \cite{hawpag} in the case
of AdS$_4$ black holes, and by Witten \cite{wit2} in the case of
AdS$_5$ black holes.

    The  bound (\ref{Bekenstein})
resembles the  controversial universal  bound suggested
by Bekenstein for systems in flat Minkowski spacetime, except that 
Bekenstein's putative universal Minkowski  bound
contains an undefined radius.
In the AdS-Bekenstein Bound (\ref{Bekenstein}),
this radius
is taken to be that of AdS$_n$. However, for large radius
we can use the Tolman redshifting formulae (which are of course
valid only for large radius since we are using an approximate
form for the metric near the boundary), and we may just as well write
\ben
2 \pi y e \ge (n-2) s\,,
\een
where now the radius is that of $S^{n-2}$.
For clarity, we shall call this latter bound
the {\it Spherical Bekenstein Bound} or the  $S^{n-2} $-Bekenstein bound.

     Note that neither the AdS$_n$ Bekenstein bound nor the $S^{n-2}$
Bekenstein bound, like that in Bekenstein's original and rather
imprecise Minkowski-spacetime bound, contain Newton's constant or make any
specific reference to gravity.  Moreover, the AdS$_n$ Bekenstein bound
(\ref{Bekenstein}) reduces, in the Minkowski limit $l\rightarrow \infty$,
to the undemanding requirement that the energy be non-negative. 

\subsection{Non-rotating Reissner-Nordstr\"om black holes}\label{mincvsec}

    Remarkably, the Cardy-Verlinde formula is satisfied by Schwarzschild-AdS
black holes in arbitrary dimensions.  However, it is violated by 
Reissner-Nordstr\"om-AdS black holes.  Nevertheless, a simple 
minimal modification does hold, namely
\ben
{(n-2)S \over 2 \pi l}= \sqrt{E_c\bigl(2E-E_c- \Phi Q \bigr)}\,,
\label{minmod}
\een
or in Pythagorean form,
\ben
  (E-\half \Phi Q ) ^2 =
 \Bigl({(n-2)  S \over 2 \pi l} \Bigr)^2 
   + \Bigl ({\Psi \over (n-2}  + \half  \Phi Q \Bigr )^2\,.
\een
The minimally modified Bekenstein bound becomes
\ben
E \ge \half \Phi Q   + {(n-2) S \over 2 \pi l}\,,
\een
with equality if and only if
\ben
 (n-2) \Psi +\half \Phi Q =0\,,
\een
or
\ben
(n-2) T I +\half \Phi Q =0\,.
\een

   Because $\Phi Q \ge0$, we see that the AdS$_n$ Bekenstein bound holds
for Reissner-Nordstr\"om-AdS black holes.

\subsection{Rotating black holes without charge}

   As observed in \cite{cai}, the Cardy-Verlinde formula does {\it
not} hold for rotating black holes in any dimension $n\ge 4$, if one
uses the thermodynamic quantities defined with respect to a
non-rotating frame at infinity.  Remarkably, however, is was found in
\cite{cai} (see \cite{klemm1,klemm2} for earlier discussions) that in
all dimensions it does hold if one uses the quantities defined with
respect to a frame that rotates with angular velocities $-a_i/l^2$ at
infinity.  Because $E\ge E'$, one obtains an inequality,
\be
E \ge E' = {1 \over 2} E_c +
{ 1 \over 2 E_c} \Big[ { (n-2) S \over 2 \pi l } \Big]^2
\,,\label{kbound} 
\een
whence
\be
E \ge \fft{(n-2) S}{2\pi l}\,.\label{bbound}
\ee
In other words, although the conserved quantities measured with 
respect to a frame non-rotating at infinity do not satisfy the
Cardy-Verlinde formula, they do satisfy the Bekenstein bound.

   In fact, one does not need to pass to the quantities $E'$ and
${\Omega^i}'$ to establish that rotating AdS black holes satisfy the
AdS$_n$ Bekenstein bound.  From (\ref{kadsarea}), we have
\be
S= \fft{ {\cal A}_{n-2}\, m l}{4 (\prod_j \Xi_j)}\, \fft{r_+/l}{1+r_+^2/l^2}
\ee
and hence, since $x/(1+x^2)\le \ft12$, we have
\be
S \le \fft{{\cal A}_{n-2}\, m l}{4 (\prod_j\Xi_j)}\,,\label{sineq}
\ee
with equality if and only if $r_+=l$.  From results in \cite{gibperpop}, 
the Euclidean action for the $n$-dimensional Kerr-AdS black hole is 
given by
\be
I = \fft{\beta {\cal A}_{n-2}\, m}{8\pi (\prod_i \Xi_i)}\, 
\fft{l^2 - r_+^2}{l^2+r_+^2}\,,
\ee
where $\beta$ is the inverse Hawking temperature.
Thus we see that $r_+=l$ corresponds to the Hawking-Page transition, 
where the Euclidean action vanishes. 

    From  (\ref{dmassdefsodd}) and (\ref{dmassdefseven}) we have
\be
E \ge \fft{{\cal A}_{n-2}\, m (n-2)}{8\pi \, (\prod_j \Xi_j)}\,,
\label{eineq}
\ee
since $\Xi_i\le 1$, with equality if and only if the black hole is
non-rotating, i.e. if and only if all $a_i=0$.  Combining (\ref{sineq})
and (\ref{eineq}) gives the AdS$_n$ Bekenstein bound (\ref{bbound}),
with equality if and only if the black hole is non-rotating, and at
the Hawking-Page transition.

\section{Further Examples}

   From the previous work, it is natural to wonder whether a simple
modification of the Cardy-Verlinde formula, involving the use of $E'$,
and the electrostatic term $\Phi Q$, continues to hold for more
complicated black holes with, for example, more than one charge, or
the recently-constructed solutions representing rotating black holes
with one or more charge, and one or more rotation parameters.  In this
section we shall study the various cases, and find that while no universal
simple modified Cardy-Verlinde formula appears to exist that covers all
these cases, we do find in many cases we have studied that 
\be
E_c'(2E'-E_c' - \Phi_i Q_i) \ge \Big(\fft{(n-2) S}{2\pi l}\Big)^2\,,
\ee
which implies the electrostatic form of the AdS$_n$ Bekenstein bound,
\be
E \ge \ft12 \Phi_i Q_i + \fft{(n-2) S}{2\pi l}\,.\label{adsbek}
\ee

\subsection{Non-rotating black holes with multiple charges}

   Modifications of the Cardy-Verlinde formula for multi-charge 
non-rotating black holes in gauged supergravities have been discussed in
\cite{cai2,klemm2}.  Here, we give a related discussion, focussing in
particular on the electrostatic AdS-Bekenstein bound, which we show
to be satisfied in all the four, five and seven-dimensional examples that
we consider.

\subsubsection{Four-dimensional multi-charge black holes}

   The general four-charge solutions in four-dimensional gauged
supergravity are given by
\bea
ds_4^2 &=& -(\prod_i H_i)^{-1/2}\, f\, dt^2 +
   (\prod_i H_i)^{1/2}\, [ f^{-1}\, d\rho^2 + \rho^2\, d\Omega_2^2 ]\,,
\nn\\
A_i &=& (1-H_i^{-1}) \sqrt{\fft{q_i+\mu}{q_i}}\, dt\,,\qquad
   X_i = (\prod_j H_j)^{1/4}\, H_i^{-1}\,,\nn\\
H_i &=& 1 + \fft{q_i}{\rho}\,,\qquad
 f = 1 - \fft{\mu}{\rho} + g^2 \rho^2 \, \prod_i H_i\,,\label{4chargesol}
\eea
where $X_i = e^{\ft12 \vec a_i\cdot \vec\varphi}$, where $\vec\varphi$ 
denotes the three canonically-normalised scalar fields, the $\vec a_i$ are
constant vectors satisfying $\vec a_i\cdot\vec a_j = 4\delta_{ij}-1$, 
and the Lagrangian is given by
\be
e^{-1} {\cal L} = \fft1{16\pi} R - \fft1{8\pi} (\del\varphi)^2 -
  \fft1{16\pi}  \sum_i X_i^{-2} F_i^2 + \fft{g^2}{16 \pi}   
                    \sum_{i<j} {X_i X_j}\,.
\ee
(We use units where $G=1$.  See \cite{tenauth} for a more extensive
discussion of the notation and conventions that we are using.)  Note
that in order for the solution to be real, and free of naked
singularities, we must have $q_i\ge0$.

   Straightforward calculations give
\bea
E &=& \ft12 \mu + \ft14 \sum_i q_i\,,\qquad
 Q_i = \sqrt{q_i\, (\mu + q_i)}\,,\nn\\
E_c &=& \ft14 \sum_i (\rho_+ + q_i)\,,\nn\\
2E-E_c-\sum_i \Phi_i Q_i &=&
\ft14 \rho_+ (\mu -\rho_+) \sum _i { 1 \over (\rho_+ + q_i
) } \,,\nn\\
S&=& \fft{\pi}{g}\,  \sqrt{\rho_+(\mu -\rho_+)}\,,
\eea
where $\rho_+$ is the radius of the horizon, i.e. the largest root
of $f(\rho)=0$.  

   One observes  that unless the charges are equal, $q_i=q$, 
the minimally modified Cardy-Verlinde
formula (\ref{minmod}) fails.  However, using the fact that
the arithmetic mean is never less than the harmonic mean,
one finds that
\ben
E-\half \sum_i Q_i \Phi_i \ge {\half E_c}  +
{g^2 S^2 \over 2 \pi^2 E_c} \,.
\een
Thus although the Cardy-Verlinde formula is violated,
\be
S \ne \fft{\pi}{g}\, \sqrt{E_c(2E-E_c- \sum_i \Phi_i Q_i)}\,,
\ee
nevertheless, the electrostatic AdS-Bekenstein bound still holds,
\ben
E  \ge \ft12 \sum_i Q_i \Phi_i +   {g S \over \pi}\,,
\een
despite the fact that the scalar fields, and hence also the
potential $-g^2 \sum_{i<j} X_i X_j$, are space dependent.  In these
cases $1/g$ 
is only the {\it asymptotic} value of the anti-de Sitter radius.

\subsubsection{Five-dimensional multi-charge black holes}

  The general three-charge solutions in five-dimensional gauged supergravity 
are given by
\bea
ds_5^2 &=& -(\prod_i H_i)^{-2/3}\, f\, dt^2 +
   (\prod_i H_i)^{1/3}\, [ f^{-1}\, d\rho^2 + \rho^2\, d\Omega_3^2 ]\,,
\nn\\
A_i &=& (1-H_i^{-1}) \sqrt{\fft{q_i+\mu}{q_i}}\, dt\,,\qquad
   X_i = (\prod_j H_j)^{1/3}\, H_i^{-1}\,,\nn\\
H_i &=& 1 + \fft{q_i}{\rho^2}\,,\qquad
 f = 1 - \fft{\mu}{\rho^2} + 
        g^2 \rho^2\, \prod_i H_i\,,\label{5chargesol}
\eea
where $X_i = e^{\ft12 \vec a_i\cdot \vec\varphi}$, where $\vec\varphi$ 
denotes the two canonically-normalised scalar fields, and the 
relevant Lagrangian is given by
\be
e^{-1} {\cal L} = \fft1{16\pi} R - \fft1{8\pi} (\del\varphi)^2 -
  \fft1{16\pi}  \sum_i X_i^{-2} F_i^2  + \fft{g^2}{4\pi} \sum_i X_i^{-1}\,.
\ee
(We again use units where $G=1$.)  Again we must have $q_i\ge 0$ in
order to have a real solution with no naked singularities.

   After straightforward calculation, we find that 
\bea
E &=& \ft38 \pi\mu + \ft14 \pi \sum_i q_i\,,\qquad 
  Q_i = \ft14 \pi \sqrt{q_i(\mu+q_i)}\,,\nn\\
E_c &=& \ft14 \pi\,  \sum_i (\rho_+^2 + q_i)\,,\nn\\
2E - E_c - \sum_i\Phi_i Q_i &=& \ft14\pi\, \rho_+^2\, 
      (\mu - \rho_+^2)\, \sum_i \fft{1}{\rho_+^2 + q_i}\,,\nn\\
S &=& \fft{\pi^2}{2g}\, \rho_+\, (\mu-\rho_+^2)^{1/2}\,,
\eea
where $\rho_+$ is the largest root of $f(\rho)=0$.  Again we see that 
the minimally-modified Cardy-Verlinde formula (\ref{minmod}) fails \
unless the $q_i$ are all equal.  Again, using the fact that the 
arithmetic mean is never less than the harmonic mean, we obtain
the inequality
\be
E_c (2E - E_c - \sum_i \Phi_i Q_i)\ge \Big(\fft{3g S}{2\pi}\Big)^2\,,
\ee
and hence we derive the minimally-modified AdS-Bekenstein bound
\be
E -\ft12\sum_i \Phi_i Q_i \ge \fft{3g S}{2\pi}\,.
\ee
Note that again, $1/g$ is only the asymptotic AdS radius.

\subsubsection{Seven-dimensional multi-charge black holes}

   The non-rotating multi-charge black-hole solutions of maximal gauged
supergravity in seven dimensions take the form
\bea
ds_7^2 &=& - (H_1 H_2)^{-4/5} \, f\, dt^2 +
   (H_1 H_2)^{1/5} \, [ f^{-1}\, d\rho^2 + \rho^2\, d\Omega_5^2 ]\,,
\nn\\
A_i &=& (1-H_i^{-1}) \sqrt{\fft{q_i+\mu}{q_i}}\, dt\,,\qquad
   X_i = (H_1 H_2)^{2/5}\, H_i^{-1}\,,\nn\\
H_i &=& 1 + \fft{q_i}{\rho^4}\,,\qquad
 f = 1 - \fft{\mu}{\rho^2} + 
        g^2 \rho^2\, H_1 H_2\,,\label{7chargesol}
\eea
with the charges carried by the $U(1)\times U(1)$ gauge fields in the 
abelian subgroup of $SO(5)$.  Straightforward calculations show that
\bea
 E &=& \ft5{16} \pi^2\mu + \ft14 \pi^2 \sum_i q_i\,,\qquad 
  Q_i = \ft14 \pi^2 \sqrt{q_i(\mu+q_i)}\,,\nn\\
E_c &=& \ft18 \pi^2\, (5 \rho_+^4 + \sum_i q_i)\,,\nn\\
2E - E_c - \sum_i\Phi_i Q_i &=& \ft18\pi^2\,\Big[ 5(\mu-\rho_+^4) + 
         3\sum_i q_i - 2 \sum_i \fft{q_i(\mu+q_i)}{\rho_+^4 + q_i}\Big]
\,,\nn\\
S &=& \fft{\pi^3}{4g}\, \rho_+^2\, (\mu-\rho_+^4)^{1/2}\,.
\eea

   The verification that these solutions satisfy the electrostatic
AdS-Bekenstein bound is slightly more complicated than for the
four-dimensional and five-dimensional cases.  This is presumably
related to the fact that unlike in $n=4$ and $n=5$, in these
seven-dimensional solutions the scalar fields are non-constant even
when the charges are set equal.  In fact the easiest way to verify the
Bekenstein bound is by performing a direct calculation of
\be
X\equiv E - \ft12 \sum_i \Phi_i Q_i - \fft{5 S g}{2\pi}\,,
\ee
and verifiying that $X$ is non-negative.  We find that 
\be
X= \fft1{\rho_+^2}\, \Big[ 5 \rho_+^6(g \rho_+ -1)^2 + q_1 q_2 g^2 +
   3(q_1+q_2) \rho_+^2 (1+g^2\rho_+^2) - 10 \sqrt{(\rho_+^4+q_1)(
\rho_+^4+q2)}\Big]\,.
\ee
Using the Maclaurin-Cauchy inequality 
\be
\prod_i b_i^{q_i} \le \sum_i b_i\, q_i\,,\qquad \hbox{where} \qquad
\sum_i q_i =1 \,,\label{maccau}
\ee
we may deduce that
\be
 \sqrt{(\rho_+^4+q_1)(\rho_+^4+q2)} \le \ft12(2\rho_+^4 + q_1 + q_2)\,,
\ee
and hence we see that
\be
X \ge  \fft1{\rho_+^2}\, \Big[ 5 \rho_+^6(g \rho_+ -1)^2 + q_1 q_2 g^2 +
   (q_1+q_2) \rho_+^2 [ 3 (g \rho_+ -\ft56)^2 + \ft{11}{12}]\Big]\,,
\ee
which proves that $X\ge0$ and hence the Bekenstein bound is satisfied.

\subsubsection{Relation to previous work}

   A more complicated modification of the Cardy-Verlinde formula has
been proposed, in terms of the parameters $q_i$ and $\mu$ which appear
in the multi-charge metrics \cite{cai2}.  This modification 
incorporates the idea that the pressure of the associated conformal 
field theory should be reduced by its electrostatic self-repulsion.
However, when re-expressed in terms of the fundamental thermodynamic variables
$E$, $\Phi_i$, $Q_i$, $T$ and $S$, the modification takes on a rather
complicated form, which appears to be a somewhat {\it ad hoc} 
construction designed to ensure the continued validity of the
Cardy-Verlinde formula in these particular examples.

   The modified Cardy-Verlinde formula in \cite{cai2} is given by 
\be
S = \fft{2\pi}{(n-2) g}\, \sqrt{\hat E_c ( 2E-2 E_q - \hat E_c)}\,,
\ee
where
\be
\hat E_c = E_c - E_q\,,\qquad E_q = \fft{ {\cal A}_{n-2}\, (n-3)}{16\pi}
   \sum_i q_i \,.
\label{ehceq}
\ee
This implies
\be
E = \ft12 \hat E_c + \fft1{2\hat E_c}\, \Big(\fft{(n-2) g S}{2\pi}\Big)^2
    + E_q\,,
\ee
and hence, minimising with respect to $\hat E_c$, one again obtains
the AdS-Bekenstein bound
\be
E\ge E_q + \fft{(n-2) g S}{2\pi}\ge  \fft{(n-2) g S}{2\pi} \,.
\ee
It is interesting to compare $E_q$, given in (\ref{ehceq}), with 
$\sum_i \Phi_i Q_i$, which is given by
\be
 \fft{ {\cal A}_{n-2}\, (n-3)}{16\pi}\sum_i 
        \fft{q_i (\mu+ q_i)}{\rho_+^{n-3} + q_i}\,,
\ee
from which it follows that $\sum \Phi_i Q_i \ge E_q$, since $\rho_+^{n-3} \le 
\mu$.

\subsection{Rotating charged black holes}

   Various solutions for charged rotating black holes in gauged
supergravities have been obtained.  In this section, we study the 
generalised Bekenstein bounds for these cases.

\subsubsection{Four-dimensional Kerr-Newman-AdS black holes}

   One might have hoped that, at least for the four-dimensional 
Kerr-Newman-AdS 
solution, a simple modification of the Cardy-Verlinde formula would work.
However, one finds that the minimally-modified Cardy-Verlinde formula,
even in in terms of of quantities measured with respect to the
canonically rotating frame, fails. Explicitly
\ben
E^\prime_c \bigl( 2E^\prime -E^\prime_c -\Phi Q   \bigr) = 
\Big(\fft{S}{\pi l}\Big)^2 +
{16 a^2 Q^2 \over l^2} \,,
\een
where $a$ is
given in terms of the extensive quantities by
\ben
a= \sqrt{{ {E^\prime}^2  l^4 \over 4 J^2} + l^2 } 
            -{E^\prime l^2 \over 2 |J|}\,.
\een

Note that
\ben
E^\prime_c \bigl( 2E^\prime -E^\prime_c -\Phi Q   \bigr)
  \ge \bigl( {S \over \pi l} \bigr)^2
\een
whence
\ben
E^\prime \ge \half \Phi Q + { \half  E_c^\prime } +
{S ^2  \over 2 \pi^2  l^2  E ^\prime_c} \ge \half \Phi Q  
  + {S \over \pi l}\,.
\een
But since $E\ge E'$, we have
\ben
E \ge \half \Phi Q  + {S \over \pi l}\,.
\een
In other words, once again, despite  the fact that 
the Cardy-Verlinde formula fails to hold,
the electrostatic AdS$_4$-Bekenstein bound still holds.

\subsubsection{Four-dimensional rotating black holes with pair-wise 
             equal charges}

    The solution four rotating black holes in four-dimensional gauged
supergravity with four charges that are set pairwise equal was given
in \cite{d4gauge}.   The thermodynamic quantities were evaluated in 
\cite{tmachine}, where it was shown that the conserved energy, angular
momentum and charges are given by
\bea
E &=& \fft{2m+q_1 + q_2}{2 \Xi^2}\,,\qquad
J = \fft{a(2m+q_1+q_2)}{2\Xi^2}\,,\nn\\
Q_1&=&Q_2= \fft{\sqrt{q_1(2m+q_1)}}{4\Xi}\,,\qquad
Q_3=Q_4= \fft{\sqrt{q_2(2m+q_2)}}{4\Xi}\,,
\eea
where the parameters $q_i$
are related to the boost parameters $\delta_i$ in \cite{d4gauge,tmachine} by
$q_i=2m \sinh^2\delta_i$. 

    We find that
\ben
E^\prime_c \bigl( 2E^\prime -E^\prime_c -\sum  _i \Phi_i  Q_i    \bigr)
                 = \bigl( {S \over \pi l} \bigr)^2 + 
    \fft{X g^2}{4\Xi\, r_+}\,,
\een
where
\bea
X&=& 2a^2 (q_1+q_2)(a^2+a^2g^2 q_1 q_2 + g^2 q_1^2 q_2^2)\nn\\
&&+ [q_1 q_2 (q_1-q_2)^2 + a^4 g^2(q_1^2 + 6 q_1 q_2 + q_2^2) \nn\\
&&\quad  + a^2((q_1-q_2)^2 + 2q_1^2 + 2 q_2^2 + 3 g^2 q_1^3 q_2 +
   3 g^2 q_2^3 q_1 + 10 g^2 q_1^2 q_2^2)] r_+ \nn\\
&& + (q_1+q_2)(2a^2 + 2 a^4 g^2 + (q_1-q_2)^2 + a^2 g^2(q_1^2+q_2^2) 
        + 10 a^2 g^2 q_1 q_2] r_+^2\nn\\
&& + [(q_1-q_2)^2 + 3 a^2 g^2 (q_1^2+q_2^2) + 10 a^2 g^2 q_1 q_2 ] r_+^3\nn\\
&& + 2 a^2 g^2 (q_1+q_2) r_+^4\,.
\eea
Since the parameters
$q_1$ and $q_2$ must be positive, the positivity of $X$ is manifest.
Hence we obtain the Bekenstein bound
\ben
E \ge \half \sum _i \Phi_i  Q_i   + {S g\over \pi}\,.
\een

\subsubsection{Five-dimensional rotating black holes with charges}

   The solution for a rotating black hole in five-dimensional minimal
gauged supergravity, with independent rotation parameters in the two
orthogonal planes in the transverse space, was obtained recently 
\cite{d5gaugeab2}.  The solution can equivalently be viewed as
a solution of $SO(6)$-gauged supergravity, with three equal charges
carried by the $U(1)^3$ abelian subgroup.  The thermodynamic quantities
were also evaluated in \cite{d5gaugeab2}, and 
it was shown that the energy, angular momenta and charge
are given in terms of the parameters $m$, $a$ and $b$ in the metric by by
\bea
E&=& \fft{\pi m(2 \Xi_a + 2 \Xi_b - \Xi_a \Xi_b) + 
          2\pi q a b g^2 (\Xi_a+\Xi_b)}{4 \Xi_a^2 \Xi_b^2}\,,\nn\\
J_a &=& \fft{2\pi m a + \pi q b (1+a^2 g^2)}{4 \Xi_a^2 \Xi_b}\,,\nn\\
J_b &=& \fft{2\pi m b + \pi q a (1+b^2 g^2)}{4 \Xi_b^2 \Xi_a}\,,\nn\\
Q &=& \fft{\pi \sqrt3 q }{4 \Xi_a \Xi_b}\,,
\eea
where $g=1/l$. We find that
\be
E_c'( 2E' - E_c' - \Phi Q) = \Big(\fft{3 S g}{2\pi}\Big)^2 +
\fft{\pi^2 g^2 q X}{16\Xi_a^2 \Xi_b^2 [(r_+^2+a^2)(r_+^2+b^2) 
                                              + a b q]\, r_+}\,,
\ee
where $X$ is given by
\bea
X &=& q^3 a^2 b^2 + q^2 ab(18 a^2 b^2 + 15 (a^2+b^2) r_+^2 + 4 a^2 b^2 r^2 
           + 9 r^4)\nn\\
&& + q (r_+^2+a^2)(r_+^2+b^2)(18 a^2 b^2 + 9(a^2+b^2) r_+^2 + 
                                10 a^2 b^2 g^2 r_+^2)\nn\\
&& +
        6 a b (r_+^2+a^2)(r_+^2+b^2)^2 (1+g^2 r_+^2)\,.
\eea
Since $X$ is manifestly positive, at least when $q$, $a$ and $b$ are positive,
we therefore obtain the Bekenstein bound
\be
E\ge \ft12 \Phi Q + \fft{3S g}{2\pi} \,.\label{d5bek}
\ee
The bound is saturated if $a=b=q=0$.

    A further solution for charged rotating five-dimensional black holes
was obtained in \cite{d5gaugeab1}, which corresponds to a case where 
the three charges carried by the $U(1)^3\in SO(6)$ gauge fields are
still all non-zero, but with two of them being equal, and the third
related to the first two in a specific way.  Thus the solutions in
\cite{d5gaugeab1} have four independent parameters, namely the mass, the
two rotations, and a parameter characterising the charges.  For these
solutions we find
\be
E_c(2 E' - E_c - \sum_i \Phi_i Q_i) = \Big(\fft{3 S g}{2\pi}\Big)^2 
+ \fft{\pi^2 g^2 q X}{16 \Xi_a^2 \Xi_b^2 r_+^2 [(r_+^2+a^2)(r_+^2+b^2)
     + q r_+^2]}\,,
\ee
where $X$ is given by
\bea
X &=& 3 (a^2+b^2)(r_+^2+a^2)(r_+^2+b^2)(1+g^2 r_+^2) +
      q^3 r_+^2 (3 a^2 b^2 g^2 + 2 r_+^2 + (a^2+b^2) g^2 r_+^2)\nn\\
&& \!\!\!\!\!\!\!\! 
   + q(r_+^2+a^2)(r_+^2+b^2)[3 a^2 b^2 + (a^2+b^2)r_+^2 (8+ 7 g^2 r_+^2)
 + g^2 r_+^2(a^4+b^4+11 a^2 b^2) + 2 r_+^4]\nn\\
&& + q^2 r_+^2 [5 a^2 b^2 + (a^2+b^2)(7 r_+^2 + 4 a^2 b^2g^2  + 5 g^2 r_+^4)  
         + 2 g^2 r_+^2(a^4+b^4+8 a^2 b^2)]\,.
\eea
This is manifestly positive when $q$ is positive, and so again the 
AdS-Bekenstein bound (\ref{d5bek}) is satisfied.

\section{The Cosmic Censorship Bound}

   So far, we have established that for many of the stationary black hole
solutions we have examined, the electrostatic AdS$_n$ Bekenstein bound
(\ref{adsbek}) is satisfied.  In this section, we shall propose that this
is a consequence of a more basic and more general lower bound for the 
energy $E$ of any initial data set for the Einstein equations with 
negative cosmological constant, coupled to a matter system that satisfies
the dominant energy condition.  This {\it Cosmic Censorship Bound} 
is expressed in terms of the area $A$ of the outermost 
apparent horizon of that initial data set.  In the case that there is no 
charge, the postulated lower bound reads
\be
E \ge \fft{(n-2) A}{16\pi l}\, \Big[ 
     l\, \Big(\fft{A}{ {\cal A}_{n-2} }\Big)^{-\ft1{n-2}} +
\fft1{l}\,  \Big(\fft{A}{{\cal A}_{n-2} }\Big)^{\ft1{n-2}} \Big]\,.
\label{ccbound}
\ee

   Some consequences of this bound, which is a more global extension of
Hawking's variational principle for black holes \cite{hawking2}, are:

\begin{itemize}

\item[(1)] If $l\rightarrow \infty$, then (\ref{ccbound}) reduces to a
bound first proposed in $n=4$ dimensions by Penrose, who observed that it is a
necessary condition for the cosmic censorship hypothesis \cite{penrose}.
(See \cite{brachr,bray}.)

\item[(2)] The proposed bound (\ref{ccbound}) implies a generalisation
of the AdS$_n$ Bekenstein bound, to the non-stationary case.  Noting
that the quantity in square brackets in (\ref{ccbound}) must be
greater than or equal to 2, we have a generalisation of Bekenstein's
bound to the time-dependent case when one may no longer equate entropy
with $1/4$ of the area of an apparent horizon:
\be
E\ge \fft{(n-2) A}{8\pi l}\,.
\ee

\item[(3)] The cosmic censorship bound (\ref{ccbound}) is attained for the
case of a Schwarzschild-anti-de Sitter black hole, and we propose that
this is the only case for which it is saturated.

\end{itemize}

    The strongest physical argument in favour of (\ref{ccbound}) is as
follows.  Consider an initial data set with total energy $E_{\rm
initial}$, a single outermost apparent horizon of area $A_{\rm
initial}$, and vanishing total angular momentum and charge.  According
to standard lore, this should settle down to a stationary state
described by a Schwarzschild-de Sitter black hole with total energy
$E_{\rm final}$ and event-horizon area $A_{\rm final}$, where
\be
E_{\rm final} = \fft{(n-2) A_{\rm final}}{16\pi l}\, \Big[ 
     l\, \Big(\fft{A_{\rm final}}{ {\cal A}_{n-2} }\Big)^{-\ft1{n-2}} +
\fft1{l}\,  \Big(\fft{A_{\rm final}}{{\cal A}_{n-2} }\Big)^{\ft1{n-2}} 
\Big]\,. \label{ccf}
\ee
Assuming cosmic censorhip, the apparent horizon
lies inside the event horizon, and since, in the time symmetric case,
the apparent horizon is a  minimal surface,
its area gives a lower bound for the area of the event horzion.
Now applying Hawking's theorem stating that the area of the
horion is non-decreasing, we obtain
\be
A_{\rm final} \ge A_{\rm initial}\,,
\ee
In anti-de Sitter spacetime, unlike in asymptotically-flat spacetimes, the
total energy is constant, and therefore 
\be
E_{\rm final} = E_{\rm initial}\,.
\ee
It follows that 
\be
E_{\rm initial} \ge \fft{(n-2) A_{\rm initial}}{16\pi l}\, \Big[ 
     l\, \Big(\fft{A_{\rm initial}}{ {\cal A}_{n-2} }\Big)^{-\ft1{n-2}} +
\fft1{l}\,  \Big(\fft{A_{\rm initial}}{{\cal A}_{n-2} }\Big)^{\ft1{n-2}} 
\Big]\,. \label{cci}
\ee

   If the initial value set had non-vanishing charge or angular
momentum, it would be expected to settle down to the relevant
stationary solution carrying those charges or angular momenta.
However, he energy due to the charge or angular momentum could be extracted 
by dropping particles carrying charge or angular momentum
into the black hole.  In this process, the area of the event horizon 
cannot decrease,  but the energy may.  Thus we expect the energy of the
black hole with charge or angular momentum to be less than that of a 
Schwarzschild-AdS black hole with the same event-horizon area.

    We shall now review some of the additional evidence for this form of
the cosmic censorship bound, and provide some further support for it.

   In the case of $n=4$ dimensions and time-symmetric initial data,
Jang and Wald's extension \cite{janwal} of Geroch's \cite{geroch} 
suggested method
of proof of the positive mass theorem for asymptotically-flat metrics
using the inverse mean-curvature flow may be extended
\cite{bougibhor,gibbons2} to cover the case of asymptotically anti-de
Sitter metrics.  Furthermore, if one does so one obtains precisely the
proposed lower bound (\ref{ccbound}).  The Geroch-Jang-Wald proposed
method of proof has been made into a rigorous theorem by Huisken and
Ilmanen \cite{huiilm1,huiilm2}. It seems plausible, but there is as yet no
rigorous proof, that their methods will extend to the anti-de Sitter
case.

   There is no general proof of the original asymptotically-flat
cosmic censorship inequality in higher dimensions.  In \cite{bafrle},
it is shown to hold in the case of a collapsing shell, using the
obvious generalisation of the four-dimensional calculations in
\cite{gibbons1}.

    There exists a natural generalisation of the inverse mean-curvature flow
to higher-dimensional time-symmetric initial-value sets \cite{gibbons2}.   
This might yield a proof of the higher-dimensional inequality if on each
level surface
\be
\int[\bar R - (n-2)(n-3) ]dA \ge0\,,
\ee
where $\bar R$ is the Ricci scalar of the $(n-2)$-dimensional metric on the
level surface, and the integration is over this surface.

   In the static spherically-symmetric case, the inequality (\ref{ccbound})
may be proved as follows.  We write the metric as
\be
ds^2 = - e^{2\nu(r)}\, \Big(1 - \fft{2{\frak{m}}(r)}{r^{n-3}}\Big)\, dt^2 +
  \Big(1 - \fft{2 {\frak{m}}(r)}{r^{n-3}}\Big)^{-1}\, 
dr^2 + r^2 d\Omega_{n-2}^2\,.\label{sphsym}
\ee
There is an horizon of area $A= {\cal A}_{n-2}\, r_+^{n-2}$ at $r=r_+$,
where 
\be
r_+^{n-3} = 2{\frak{m}}(r_+)\,.
\ee
The Einstein equations 
\be
R_{\mu\nu} - \ft12 R\, g_{\mu\nu} = 8\pi T_{\mu\nu} 
\ee
(where we include the contribution of the cosmological constant in
$T_{\mu\nu}$) imply
\be
\fft{d{\frak{m}}}{dr} = \fft{8\pi}{n-2}\, r^{n-2}\, T_{\hat t\hat t}\,,
\label{dmdr}
\ee
where the hats indicate components in an orthonormal frame.  We have 
\be
T_{\mu\nu} = T_{\mu\nu}^{\rm cosmic }  + T_{\mu\nu}^{\rm matter}\,,
\ee
where $T_{\mu\nu}^{\rm cosmic}$ is the contribution from the
cosmological term.   As $r$ tends to infinity,
\be
{\frak{m}}(r)\longrightarrow m - \fft{r^{n-1}}{2 l^2}\,,
\ee
where $l$ is the asymptotic de Sitter radius.  We may integrate
(\ref{dmdr}) from the horizon to infinity, to obtain
\be
 m = \ft12 r_+^{n-3}  + \fft{r_+^{n-1}}{2l^2} + \fft{8\pi}{n-2}\, 
      \int_{r_+}^\infty dr\, r^{n-2}\,  T_{\hat0\hat0}^{\rm matter}
+\int_{r_+}^\infty dr r^{n-2}\, \Big[ \fft{n-1}{2 l^2} + \fft{8\pi}{n-2}\, 
   T_{\hat t\hat t}^{\rm cosmic} \Big]\,.
\ee
If $T_{\hat t\hat t}^{\rm cosmic}$ is constant and 
$T_{\hat t\hat t}^{\rm matter}$ satisfies the positive-energy
condition, we obtain the cosmic censorship bound, which will be 
saturated if and only if $T_{\hat t\hat t}^{\rm matter}=0$, i.e. for
the Schwarzschild-de Sitter metric.

   If $T_{\hat t\hat t}^{\rm cosmic}$ is not constant, because of the 
presence of varying scalar fields, we still obtain a lower bound for
$m$ if the integrand is positive.  Unfortunately, in the gauged 
supergravities we have considered the integrand is in fact negative, 
because the potential is in general more negative than its negative value
at its vanishing-scalar stationary point.  However, even in this case the
cosmic censorship bound may continue to hold, because kinetic energy
term for the scalars is positive.  Indeed, this is what happens in the
examples we have examined.

\subsection{Cosmic censorship for Kerr-AdS black holes in 
arbitrary dimension}

   It is straightforward to show that the general Kerr-AdS black holes in
arbitrary spacetime dimension $n$ satisfy the cosmic censorship bound
(\ref{ccbound}).  First, we note that one can characterise the 
Kerr-AdS metrics by their rotation parameters $a_i$, together with the
radius $r_+$ of the outer horizon.  The mass parameter $m$ appearing in
the metric (\ref{bl}) is then solved for using $V(r_+)=m$.   
It is then helpful to introduce a new parameterisation in terms of 
$y_i$ and $z$ instead of $a_i$ and $r_+$, where  
\be
y_i \equiv \fft{r_+^2 + a_i^2}{r_+^2\,  \Xi_i}\,,\qquad
   z= \fft{r_+}{l}\,.\label{yizdef}
\ee
Clearly we must have have $y_i\ge 1$, $z\ge 0$, and  
\be
r_+ = z l\,,\qquad a_i^2 = \fft{(y_i-1) z^2 l}{(1+z^2 y_i)}\,,\qquad
\Xi_i = \fft{1+z^2}{1+z^2 y_i}  \,.\label{yiz2}
\ee

     We begin by considering the case when $n=2N+1$ is odd.  From
(\ref{uvw}), (\ref{dmassdefsodd}) and (\ref{kadsarea}), the energy $E$
is given by
\be
E = \fft{r_+^{n-3}\,{\cal A}_{n-2}\,  
 (\prod_j y_j)}{16\pi}\Big( n-2 + 2 z^2\, \sum_i y_i \Big)\,,
\ee
whilst the right-hand side of (\ref{ccbound}) is given by
\be
\fft{ (n-2) r_+^{n-2}\,(\prod_j y_j)}{16\pi\, l}\, \Big[
    z (\prod_i y_i)^{\ft1{n-2}} + \fft1{z}\, 
              (\prod_i y_i)^{-\ft1{n-2}}\Big]\,,
\ee
and so to show that the cosmic censorship bound is satisfied, we must show
that
\be
(n-2)\Big[ 1-(\prod_i y_i)^{-\ft1{n-2}} \Big] + 
            z^2\, \Big[2\sum_i y_i -1 -(n-2)
  (\prod_i y_i)^{\ft1{n-2}} \Big]\ge 0\,.\label{ccineq1}
\ee

   The first bracketed term in (\ref{ccineq1}), i.e. the term independent of
$z$, is manifestly positive since $y_i\ge 1$.  For the terms at order $z^2$
we may use the Maclaurin-Cauchy inequality (\ref{maccau}) 
to show that
\be
(\prod_i y_i)^{\ft1{n-2}} \le \fft{1}{n-2} \, \sum_i y_i + \fft{n-3}{2(n-2)}\,.
\ee
Substituting this into the second bracketed term in (\ref{ccineq1}) shows that 
\be
2\sum_i y_i -1 -(n-2) (\prod_i y_i)^{\ft1{n-2}} \ge 
 \sum_{i=1}^N (y_i-1) \ge0\,,
\ee
and hence the cosmic censorship bound is proved.

   In the case of even dimensions $n=2N+2$, an analogous calculation shows
that cosmic censorship is satisfied if
\be
(n-2)\Big[1 - (\prod_i y_i)^{-\ft1{n-2}}\Big] + z^2\, \Big[2\sum_i y_i -
   (n-2) (\prod_i y_i)^{\ft1{n-2}}\Big]\ge0\,.\label{ccineq2}
\ee
Again using (\ref{maccau}), we can show that
\be
 (\prod_i y_i)^{\ft1{n-2}} \le \fft1{n-2} \sum_i y_i + \ft12
\ee
and so since $y_i\ge 1$, the inequality (\ref{ccineq2}) can indeed be 
seen to hold.

\subsection{Cosmic censorship for four-dimensional charged rotating
black holes}

  The cosmic censorship bound (\ref{ccbound}) can be generalised in
the case of charged black-hole solutions \cite{jang,gibbons2}.  
In four dimensions, it becomes
\be
E \ge \fft{A}{8\pi l}\, \Big[ l \Big(\fft{A}{4\pi}\Big)^{-1/2}
 + \fft1{l}\, \Big(\fft{A}{4\pi}\Big)^{1/2} + l Q^2\, 
   \Big(\fft{A}{4\pi}\Big)^{-3/2}\Big]\,,\label{ccq4}
\ee
with equality being attained for the Reissner-Nordstr\"om-AdS solution.
Calculating $E^2$ minus the square of the right-hand side of 
(\ref{ccq4}) for the Kerr-Newman-AdS solution, we find
\be
E^2 - (\rm{RHS})^2 = \fft{4 (1+g^2 r_+^2)[a^2 + q^2 + r_+^2 + 
      g^2 r_+^2 (r_+^2+a^2)]^2}{\Xi^4 r_+^2 (r_+^2 + a^2)}\,,
\ee
which is manifestly positive, thus demonstrating that the inequality
(\ref{ccq4}) is obeyed in this case.

\subsection{Cosmic censorship for five-dimensional charged
rotating black holes}

   For solutions of five-dimensional minimal gauged supergravity,
the generalised cosmic censorship bound can be written as
\be
\fft{8E}{3\pi} \ge \Big(\fft{A}{2\pi^2}\Big)^{\ft23} +
     g^2  \Big(\fft{A}{2\pi^2}\Big)^{\ft43} +
   \fft{16 Q^2}{3\pi^2}\,  \Big(\fft{A}{2\pi^2}\Big)^{-\ft23}\,,
\label{d5chargecc}
\ee
with equality being attained in the case of the non-rotating 
Reissner-Nordstr\"om-AdS black hole.  From the results obtained in
\cite{d5gaugeab2}, we find that the inequality (\ref{d5chargecc}) 
translates into the requirement that
\bea
&&\ft23 h[ 3 + z^2 (y_1+y_2)] + \ft13 y_1 y_2 [3-z^2 + 2z^2 (y_1+y_2)] 
\nn\\
&& +  \fft{h^2(1+z^2)}{3(y_1-1)(y_2-1)} [3-z^2 + 2 z^2(y_1+y_2)] 
  - (y_1 y_2 + h)^{\ft23} - z^2 (y_1 y_2 + h)^{\ft43}\nn\\
&& 
   -\fft{h^2 (y_1 y_2 + h)^{-2/3}}{(y_1-1)(y_2-1)} \, 
       (1+z^2 y_1)(1+z^2 y_2)\ge0\,,\label{hineq}
\eea
where 
\be
y_1= \fft{r_+^2+a^2}{r_+^2 \, \Xi_a}\,,\qquad
y_2= \fft{r_+^2+b^2}{r_+^2 \, \Xi_b}\,,\qquad z= g r_+\,,\qquad 
   h = \fft{ab q}{r_+^4\, \Xi_a\, \Xi_b}\,,
\ee
with $y_i\ge 1$, $z> 0$, $h\ge 0$. We have studied (\ref{hineq})
numerically and find that it appears to be satisfied for all allowed
values of the parameters $(y_1,y_2,z,h)$, and thus it appears that the
generalised cosmic censorship bound is obeyed by the five-dimensional
charged rotating AdS black holes obtained in \cite{d5gaugeab2}.

\section{An Upper Bound for the Temperature?}

   It is well known that the presence of matter with a positive energy
density tends to reduce the temperature of a black hole, because of
the redshift produced by the gravitational field of the matter.  It is
also well known that charged or rotating black holes tend to have a
smaller temperature for the same entropy than their neutral or
non-rotating versions.  A general explanation for this observation was
provided by Visser in the static spherically-symmetric case in four
dimensions with no cosmological term \cite{visser}. In this section we shall
generalise Visser's observation, and apply it to the
Hawking-Page transition.

    For the spherically-symmetric static metric (\ref{sphsym}), the Einstein 
equations imply
\bea
{d{\frak{m}} \over dr}&=& \fft{8 \pi r^{n-2}\,  T_{\hat t\hat t}}{(n-2)}\,\\
{d\nu  \over dr}&=& \fft{8 \pi r^{n-2} \, ( T_{\hat t \hat t} 
               + T_{\hat r \hat r})}{(n-2)[r^{n-3} - 2{\frak{m}}(r)]}\,,
\eea
where the  hats indicate components  in an orthonormal frame.

    The surface gravity is
\ben
\kappa= 2 \pi T = {1 \over 2 r_+ } e^{\nu(r_+)}  
    \Big( n-3 -2 r^{4-n}\, \fft{d{\frak{m}}}{dr}\Big))\,.
\een
This becomes
\be
\kappa= 2 \pi T = { 1 \over 2 r_+ } e^{\nu(r_+)}  
   \Big[(n-3)-\fft{16\pi r_+^2}{(n-2)}\, (  T_{\hat t\hat t}^{\rm cosmic} + 
        T_{\hat t\hat t}^{\rm matter}) \Big]\,.\label{kapp2}
\ee
If $T_{\hat t\hat t}^{\rm cosmic}$ is constant, then (\ref{kapp2}) becomes
\be
\kappa= 2 \pi T = { 1 \over 2 r_+ } e^{\nu(r_+)}  
    \Big[(n-3)  + (n-1) g^2 r_+^2   - 
                    \fft{16\pi r_+^2}{(n-2)}\, 
         T_{\hat t\hat t}^{\rm matter}\Big]\,.\label{kapp3}
\ee
If the matter satisfies the dominant energy condition then 
\ben
T_{\hat t\hat t}^{\rm matter} \ge |T_{\hat r \hat r}^{\rm matter} | \ge 0\,.
\een
Moreover 
\ben
\nu(r)= - \int_r ^\infty  { 8 \pi {r'}^{n-2}\, 
\bigl(T_{\hat t \hat t}^{\rm matter}
 + T_{\hat r \hat r}^{\rm matter} \bigl)
 \over 
               (n-2)[{r'}^{n-3} -2 {\frak{m}}(r') ]}\, dr' \,.
\een
Thus $\nu(r)$ will be non-positive and
\ben \fbox{$\displaystyle
4 \pi T  \le \fft{(n-3)}{ r_+}  + (n-1) g^2 r_+\,. $}\label{tineq}
\een
If $T_{\hat t\hat t}^{\rm cosmic}$ is not constant, one might expect
the kinetic term for the scalars to compensate for any extra positive 
contribution from $-T_{\hat t\hat t}^{\rm cosmic}$, and the inequality
to continue to hold.

   In the Schwarzschild-AdS case the inequality (\ref{tineq})
becomes an equality.  The minimum value of the right-hand side occurs
at
\be
r_+= \fft{1}{g}\, \sqrt{\fft{n-3}{n-1}}\,,
\ee
at which
\be
T = \fft{g \sqrt{(n-1)(n-3)}}{2\pi}\,.\label{mint}
\ee
This lower bound for the temperature is associated with the
Hawking-Page phase transition.  Below this temperature, there is no
black-hole solution, whilst above it, there are two.  The Hawking-Page
transition itself occurs at $r_+=1/g$, for which $T= (n-2)g/(2\pi)$.  For
temperatures greater than (\ref{mint}) but smaller than than
$(n-2)g/(2\pi)$, both Schwarzschild-AdS solutions have larger Euclidean
action than that of anti-de Sitter spacetime.

     The general inequality (\ref{tineq}) for spherically-symmetric
black holes may be recast in the form
\be
4\pi T\le  (n-3) \Big(\fft{A}{{\cal A}_{n-2} }\Big)^{-\ft1{n-2}} + 
                 (n-1) g^2 \Big(\fft{A}{{\cal A}_{n-2} }\Big)^{\ft1{n-2}}\,,
\label{tinequal}
\ee
where $A$ is the area of the outer horizon.  Equality is achieved in the
case of Schwarzschild-AdS black holes.  

   One might think that when the inequality is expressed in 
the form (\ref{tinequal}), it would continue to hold for rotating as
well as non-rotating black holes.  In other words, one might conjecture  
that the minimum temperature, as a function of
entropy,  is always less than or equal to the minimum temperature of
the Schwarzschild-AdS case.  In fact, we find that the bound 
(\ref{tinequal}) is obeyed bay all Kerr-AdS black holes in $n=4$ and 
$n=5$ diemnsions.   However, counterexamples
can be found for Kerr-AdS black holes in all
dimensions greater than or equal to 6.  

      To discuss the situation in arbitrary dimensions, it is
again helpful to use the parameterisation introduced in
(\ref{yizdef}).  In odd dimensions $n=2N+1$, showing that the inequality
(\ref{tinequal}) is obeyed is equivalent to showing that
\be
(n-3) (\prod_i y_i)^{-\ft1{n-2}}
   - 2 \sum_i y_i^{-1} +2  + 
   (n-1) [(\prod_i y_i)^{\ft1{n-2}}- 1]\, z^2 \ge0\,.\label{oddz2}
\ee
This must hold for the $z^0$ and $z^2$ terms independently.  It is 
clearly true for the $z^2$ terms, since $y_i\ge 1$, and so checking the
temperature bound for odd-dimensional Kerr-AdS black holes amounts to
checking whether
\be
 (N-1) (\prod_{i=1}^N y_i)^{-\ft1{2N-1}}  \ge \sum_{i=1}^N  y_i^{-1} -1
\label{oddbound}
\ee
for all $y_i\ge 1$. It is straightforward to see that in five dimensions,
for which $N=2$, the function
\be
(y_1 y_2)^{-1/3} - \fft1{y_1} - \fft1{y_2} + 1
\ee
is non-negative for all $y_i\ge 1$, since it can be written in the 
manifestly  non-negative form
\be
(y_1 y_2)^{-1}\, \Big\{(y_1-1)(y_2-1) + [(y_1 y_2)^{2/3} - 1]\Big\}\,.
\ee
This shows that all Kerr-AdS black holes in 
five dimensions obey the temperature bound (\ref{tinequal}).  
However, if $N\ge3$ it is clear that the inequality in (\ref{oddbound})
can be violated for valid choices of the parameters $y_i$.  For example,
we can take $y_1=y_2=1$, thus ensuring that the right-hand side of 
(\ref{oddbound}) is at least 1, and then choose the remaining $y_i$ large
enough so that the left-hand side of (\ref{oddbound}) is less than 1. 
Clearly if $z$, which is independently specifiable and subject only to the
restriction $z\ge 0$, is chosen to be
sufficiently small, then the order $z^2$ terms in (\ref{oddz2}) will not
be sufficiently positive to overwhelm the negative contribution from 
the terms at order $z^0$, and so (\ref{oddz2}) will be violated.  

   In even dimensions 
$n=2N+2$, the analogous calculation shows that for these Kerr-AdS black
holes the temperature inequality (\ref{tinequal}) is equivalent to
\be
(n-3) (\prod_i y_i)^{-\ft1{n-2}}
   - 2 \sum_i y_i^{-1} + 1  + 
   (n-1) [(\prod_i y_i)^{\ft1{n-2}}- 1] \, z^2\ge0\,.\label{evenz2}
\ee
Again, the terms at order $z^2$ are clearly positive, and so showing that
(\ref{evenz2}) is satisfied is equaivalent to showing that 
\be
 (N-\ft12) (\prod_{i=1}^N y_i)^{-\ft1{2N}}  \ge  
      \sum_{i=1}^N y_i^{-1} - \ft12\,.
\label{evenineq}
\ee
Clearly this inequality is always obeyed in four dimensions, corresponding
to $N=1$, since the function
\be
\ft12 y_1^{-1/2} - \fft1{y_1} + \ft12 = \ft12 y_1^{-1}\, [
  (y_1-1) + (y_1^{1/2} -1)]
\ee
is manifestly non-negative for all $y_1\ge 1$.  Thus all Kerr-AdS
black holes in four dimensions obey the temperature bound
(\ref{tinequal}).  It is clear, however, that the inequality
(\ref{evenineq}) can be violated for valid choices of the parameters,
$y_i\ge1$, if $N$ is greater than or equal to 2 (i.e. in even
dimensions $n$ greater than or equal to 6).  For example, we could
take $y_1=1$, and then by taking the remaining $y_i$ large enough, the
left-hand side of (\ref{evenineq}) can be made arbitrarily small,
while the right-hand side exceeds $\ft12$.  By also taking $z$
sufficiently small, this means that (\ref{evenz2}) can be violated
when $N\ge 2$.

      More generally, one can see that in all dimensions $n\ge 6$, there
exist regions in the $(y_i,z)$ parameter space for which the inequalities
(\ref{oddz2}) or (\ref{evenz2}) are violated, and using (\ref{yiz2}) these
can be translated back into regions in the parameter space for $(a_i, r_+)$
for which the temperature inequality (\ref{tinequal}) is not obeyed.

   It is also worth remarking that similar conclusions are obtained
if we consider asymptotically flat, rather than asymptotically AdS,
rotating black holes.  From (\ref{yizdef}) we see that the asymptotically
flat case, which arises when $g=0$, corresponds to taking $z$ to zero.
We saw above that in the asymptotically AdS case there were terms in 
the inequality that were of order $z^2$, and terms of order $z^0$.  The
former were always consistent with the inequality, and it was the 
$z^0$ terms, which are the ones that survive in the $g\rightarrow0$ limit, 
that had to be investigated in more detail.  Thus the conclusions for
asymptotically-flat rotating black holes are the same as those for
asymptoticallty-AdS rotating black holes, namely that violations of
the temperature inequality (\ref{tinequal}) can occur in all dimensions
6 and higher, in cases where some of the rotations are small and some are
large.

  Finally, we should emphasise that our finding of violations of the
inequality (\ref{tinequal}) does not contradict or threaten any cherished
beliefs.  The inequality was derived for static solutions, and,
although commonly such considerations can lead to conjectured inequalities
that have a wider range of applicability, as in the case of the cosmic
censorship bound (\ref{ccbound}), there is no {\it a priori} reason
why it should do so in this case.  The result could, perhaps, be viewed as
a salutary reminder that a conjecture that holds up well in low
dimensions may run into trouble in higher dimensions.

\section{Cosmological Event Horizons}

    In this section we take the cosmological constant to be positive,
thus
\ben
R_{\mu \nu}= {n-1 \over l^2}g_{\mu\nu}\,.
\een  
In order to obtain the necessary formulae one makes the substitution
$l^2 \rightarrow -l^2 $.

    In the case of pure de Sitter spacetime, $dS_n$,
one has  $m=a_i=0$ and  there is a cosmological horizon \cite{gibhaw}
at $r=l$. If $m>0$, this is at
\ben r=r_C \le l \,.
\een
Inside the cosmological horizon  there will, in general, be a black hole 
horizon, at $r=r_H$ say. 

    If the spacetime dimension $n$ is even, then the area of the
cosmological horizon $A_C$ is easily seen to be bounded above by the
value in pure $dS_n$,
\ben
A_C \le {\cal A}_{n-2} l^{n-2}\,.
\een
(For the four-dimensional case, see \cite{bougibhor,shnakoma}.)
In some sense, $S_{\rm max}={1 \over 4} {\cal A}_{n-2}l^{n-2} $
represents the largest amount of information
that can ever be lost through the cosmological horizon.

    By manipulations similar to those in the case of a negative
cosmological constant, one may convince oneself that
\ben
\bigl ( { A _C \over {\cal A}_{n-2} } )^{n-3 \over n-2}
 \Bigl ( 1- { 1 \over l^2 }
\bigl ( {A _C \over  {\cal A}_{n-2} }  )^{2 \over n-2} \Bigr)
\le
\bigl ({ A _H \over {\cal A}_{n-2}}  )^{n-3 \over n-2}
 \Bigl (1- { 1 \over l^2 }
\bigl ({ A _H \over  { \cal A}_{n-2}}  )^{ 2 \over n-2} \Bigr)
\een
with equality only for the  Kottler, i.e. Schwarzschild-de Sitter, solution.
In the Reissner-Nordstr\"om-de Sitter case, one can do more, and obtain
\ben
\bigl ( { A _C \over {\cal A}_{n-2} } )^{n-3 \over n-2}
 \Bigl ( 1- { 1 \over l^2 }
\bigl ( {A _C \over  {\cal A}_{n-2} }  )^{2 \over n-2} \Bigr)-{\Phi_C Q}
\le
\bigl ({ A _H \over {\cal A}_{n-2}}  )^{n-3 \over n-2}
 \Bigl (1- { 1 \over l^2 }
\bigl ({ A _H \over  { \cal A}_{n-2}}  )^{ 2 \over n-2} \Bigr) -\Phi_H Q\,,
\een
where $\Phi_C$ and $\Phi_H$ are the electrostatic potentials of the
cosmological horizon and black hole horizon.  Actually, only the
potential difference between the two horizons enters the inequality,
as must be the case by gauge invariance.

    An interesting question is whether there is an upper bound to the
area of a black hole in a background de Sitter spacetime
\cite{hashna}.  For the Schwarzschild-de Sitter solution, there is
such an upper bound, which occurs when the two horizons coincide.
This happens when
\ben
r_C=r_H= l \sqrt{n-3 \over n-1}\,.
\een
It is natural therefore to conjecture that more generally,
\ben
A_H \le {\cal A}_{n-2} l^{n-2}
\Big( {n-3 \over n-1 } \Big)^{\ft12(n-2)}\,.\label{ahineq}
\een
It is easy to check that this is true
for the Reissner-Nordstr\"om-de Sitter solution in any dimension.
The radius $r$  at which the two horizons coincide is easily seen
to be less than $l \sqrt{n-3 \over n-1}$ and so the
area of a charged black hole in a background de Sitter spacetime is indeed
never greater than
$ {\cal A}_{n-1} l^{n-2} \bigl ( {n-3 \over n-1 } \bigr )^{n-2}  $.

   In the rotating case the black hole and cosmological horizons
coincide when $m$, considered as a function of $r$, has a vanishing
derivative.  One may check that this happens at
\ben
r_C= r_H <  l \sqrt{n-3 \over n-1}\,.
\een
We have verified that the inequality (\ref{ahineq}) is satisfied for 
rotating black holes in a variety of cases.  These include the general
Kerr-de Sitter metrics in four and six dimensions; the Kerr-de Sitter
metrics with equal angular momenta in five and seven dimensions, and the
Kerr-de Sitter metrics with equal angular momenta in all even dimensions.

   Here, we shall just present the proof for the case of Kerr-de
Sitter metrics with equal angular momenta in all even dimensions
$n=2N+2$.  From the formulae collected in the appendix, and setting
$a_i=a$, we can show that the condition for double root $r_C=r_H$ can
be expressed as
\be
l^2 = \fft{r_H^2 [(n-1) r_H^2 - a^2]}{(n-3) r_H^2 - a^2}\,,
\ee
whilst the area of the horizon is given by
\be
A_H = {\cal A}_{n-2}\,  l^{n-2}\, 
   \Big( \fft{r_H^2+a^2}{l^2+a^2}\Big)^{\ft{n-2}{2}}\,.
\ee
It is straightforward to see that
\be
\fft{r_H^2+a^2}{l^2+a^2} = \fft{n-3}{n-1} - 
           \fft{2a^2}{(n-1)[ (n-1)r_H^2 -a^2]} \le  \fft{n-3}{n-1} \,,
\ee
thus proving that these Kerr-AdS black holes indeed satisfy the
bound (\ref{ahineq}).

\section{Conclusions}  

   In this paper, we have studied the relation between the
thermodynamics of the bulk variables describing rotating black holes
in gauged supergravities, and the corresponding variables in the
boundary CFT.   We have shown that by using the standard UV/IR
connection between the bulk and the boundary, bulk quantities that
satisfy the first law of thermodynamics are mapped into boundary
quantities that likewise satisfy the first law of thermodynamics.  An 
important point when considering rotating AdS black holes is that the
natural conformal boundary at large distance is defined with respect
to a coordinate frame in which the metric is asymptotically static and
asymptotically spherical.

   Our results have clarified some previous puzzling claims in the
literature, including in particular the assertion that to get boundary
quantities that satisfy the first law one must start from bulk
quantities that do not.  This assertion was based on calculations
performed in a specific frame that is rotating and non-spherical at
infinity, with an angular velocity that depends on the rotation parameters
of the black hole.  In our opinion this is not a convenient or natural
frame to use, and we believe that this is why it led to apparently
puzzling conclusions.

   In this context, it is perhaps worth remarking that in much of the
literature on the subject of rotating AdS black holes, there is a
tendency to refer to just two choices of frame, namely the frame that
is asymptotically static, and the frame with rotation rates given by
(\ref{ominf}) at infinity.  In our opinion, the discussion of whether
the thermodynamic quantities such as energy should be defined with
respect to the former or the latter frame is misplaced.  In reality
there are infinitely many different frames that could be chosen, with
arbitrary choices of asymptotic rotation rates.  Asymptotically static
frames enjoy a preferred status, and, as we showed in
\cite{gibperpop}, the quantities defined in an asymptotically static
frame satisfy the first law of thermodynamics. Frames whose asymptotic
rotation rates depend upon the black-hole rotation parameters (such as
the frame specified by (\ref{ominf})) seem to be particularly
unnatural from the point of view of thermodynamic discussions, since
one would need to include extra terms to compensate for the changing
centrifugal and Coriolis contributions to the energy.  Furthermore, if
physical results (such as the energy) depend upon the choice of frame
(in the sense that they depend upon the choice of timelike Killing
vector used to define the energy, etc.), then a justification is
called for as to why some specific frame, rather than one with some
other rate of rotation, has been chosen.  We have argued that for
thermodynamic discussions, at least, the asymptotically static frame
is the physically natural one.  

    Some of the results in \cite{klemm2,cai} show that in a different
context, namely the discussion of the Cardy-Verlinde formula, there is
a significant merit to considering the energy function $E'$ defined
with respect to the frame with asymptotic angular velocity given by
(\ref{ominf}).  It was shown in five dimensions in \cite{klemm2}, and
in higher dimensions in \cite{cai}, that the Cardy-Verlinde formula
(\ref{CardyVerlinde2}) holds for rotating AdS black holes, provided
that one uses energies and angular velocities measured with respect to
the frame with angular velocities given by (\ref{ominf}) at infinity.
As far as we are aware, there is no {\it a priori} reason why a frame 
with this particular angular velocity should be singled out in this
context, but the observation is certainly an interesting one.

   Results had also been obtained for modifications to the
Cardy-Verlinde formula when applied to non-rotating charged black
holes \cite{klemm1,klemm2,cai2}.  The recent construction of black
holes that have both rotation and charge has provided a wider spectrum
of examples where the Cardy-Verlinde formula can be tested, and we
have reported some results in the present paper.  It seems that there
is no natural and universal modification which encompasses all the
cases.  Nevertheless, as we have shown, there is a closely related
and physically more significant result that does always hold for all
the rotating charged black holes, namely the existence of an 
AdS-Bekenstein bound (\ref{Bekenstein}), and its electrostatic 
generalisation (\ref{adsbek}).  The AdS-Bekenstein bound is itself a 
consequence of a more fundamental cosmic censorship bound, and we
have explicitly demonstrated for many of the rotating and charged black 
holes that this bound is indeed satisfied.

   We have also examined the question of whether there is an upper
bound for the temperature as a function of entropy for black holes in
AdS backgrounds.  In four and five dimensions, we found that the
temperature of a rotating AdS black hole is always less than that of
the Schwarzschild-AdS black hole of the same entropy.  In six or more
dimensions, by contrast, we find that for certain choices of the
rotation parameters, the rotating AdS black hole can have a higher
temperature than the Schwarzschild-AdS black hole of the same entropy.
Finally, we discussed area inequalities for rotating black holes with
a positive cosmological constant, for which there is a cosmological
horizon as well as a black hole horizon.

\bigskip

\noindent{\Large{\bf Acknowledgements}}

    G.W.G. thanks the Centre for Mathematical Sciences, Zheijiang
University, Hangzhou, and C.N.P. thanks the Relativity and Cosmology
group, Cambridge, for hospitality during the course of this work.

\bigskip
\bigskip\bigskip

\appendix

\section{General Kerr-AdS Black Hole in Arbitrary Dimensions}
 
   In this appendix, we collect some general results on rotating
asymptotically AdS black holes in arbitrary dimension $n$.  The solution
was obtained in $n=4$ in \cite{carter}, in $n=5$ in \cite{hawhuntay}, and
in $n\ge 6$ in \cite{gilupapo1,gilupapo2}.  Results on the thermodynamics 
of the arbitrary-dimension rotating AdS black holes were obtained in
\cite{gibperpop}.  

    The metrics have $N\equiv [(n-1)/2]$ independent rotation
parameters $a_i$ in $N$ orthogonal 2-planes.  We have $n=2N+1$ when
$n$ is odd, and $n=2N+2$ when $n$ is even. Defining $\ep\equiv (n-1)$
mod 2, so that $n=2N+1+\ep$, the metrics can be described by
introducing $N$ azimuthal angles $\phi_i$, and $(N+\ep)$ ``direction
cosines'' $\mu_i$ obeying the constraint
\be
\sum_{i=1}^{N+\ep} \mu_i^2 =1\,.\label{muconstraint}
\ee
In Boyer-Lindquist type coordinates that are asymptotically non-rotating, 
the metrics are given by
\cite{gilupapo1,gilupapo2}
\bea
ds^2 &=& - W\, (1 + r^2\, l^{-2} )\, dt^2
 + \fft{2m}{U}\Bigl(W\,dt
 - \sum_{i=1}^N \fft{a_i\, \mu_i^2\, d\varphi_i}
  {\Xi_i }\Bigr)^2
 + \sum_{i=1}^N \fft{r^2 + a_i^2}{\Xi_i}\,\mu_i^2\,
    d\varphi_i^2 \nn\\
&&
 + \fft{U\, dr^2}{V-2m}
 + \sum_{i=1}^{N+\ep} \fft{r^2 + a_i^2}{\Xi_i}\, d\mu_i^2
 - \fft{l^{-2}}{W\, (1 + r^2\, l^{-2})}
    \Bigl( \sum_{i=1}^{N+\ep} \fft{r^2 + a_i^2}{\Xi_i}
    \, \mu_i\, d\mu_i\Bigr)^2 \,,\label{bl}
\eea
where
\bea
W &\equiv& \sum_{i=1}^{N+\ep} \fft{\mu_i^2}{\Xi_i}\,,\qquad
U \equiv  r^{\ep}\, \sum_{i=1}^{N+\ep} \fft{\mu_i^2}{r^2 + a_i^2}\,
\prod_{j=1}^N (r^2 + a_j^2)\,,\\
V &\equiv& r^{\ep-2}\, (1 +r^2\, l^{-2})\,
   \prod_{i=1}^N (r^2 + a_i^2)\,,\qquad \Xi_i\equiv 1 - a_i^2\, l^{-2}\,.
\label{uvw}
\eea
They satisfy $R_{\mu\nu}=-(n-1)\, l^{-2}\, g_{\mu\nu}$.

   The constant-$r$ spatial surfaces at large distance are inhomogeneously
distorted $(n-2)$-spheres.  Making the coordinate transformations
\be
\Xi_i \,  y^2 \, \hat \mu_i^2 =(r^2 +a_i^2)\, \mu_i^2\,,
\ee
where $\sum_i \hat\mu_i^2=1$, the metrics at large $y$ approach the
standard AdS form
\be
d\bar s^2 = - (1+ y^2 l^{-2}) dt^2 + \fft{dt^2}{1+ y^2 l^{-2}}
  + y^2 \sum_{k=1}^{N+\ep} (d\hat\mu_k^2 + \hat\mu_k^2 d\varphi_k^2)\,,
\ee
with round $(n-2)$-spheres of volume ${\cal A}_{n-2}\, y^{n-2}$ at
radius $y$, where ${\cal A}_{n-2}$ is the volume of the unit $(n-2)$-sphere.

   The angular velocities of the horizon, measured relative to the frame
that is non-rotating at infinity, are given by
\be
\Omega^i = \fft{(1 +r_+^2\, l^{-2})\, a_i}{r_+^2 + a_i^2}\,,\label{omapp}
\ee
and the angular momenta are
\be
J_i = \fft{m a_i {\cal A}_{n-2}}{4\pi \Xi_i\, (\prod_j \Xi_j)}\,,
\ee
where
\be
{\cal A}_{n-2}= \fft{2\pi^{(n-1)/2}}{\Gamma[(n-1)/2]}
\ee
is the volume of the unit $(n-2)$-sphere.
As shown in \cite{gibperpop}, the energy of the black hole, again measured
in the asymptotically static frame, is given by
\bea
\hbox{$n$= odd}: &&   \fbox{$\displaystyle E =
           \fft{m\, {\cal A}_{n-2}}{4\pi\, (\prod_j \Xi_j)}\,
\Big( \sum_{i=1}^N \fft1{\Xi_i} - \fft12\Big)\label{massodd} $}
\label{dmassdefsodd}\\
\hbox{$n$= even}: &&  \fbox{$\displaystyle E =
   \fft{m\, {\cal A}_{n-2}}{4\pi\, (\prod_j \Xi_j)}\,
\sum_{i=1}^N \fft1{\Xi_i}\label{masseven} $}\label{dmassdefseven}
\eea
The area of the event horizon is given by
\be
A = {\cal A}_{n-2}\, r_+^{\ep-1}\, \prod_i \fft{r_+^2 + a_i^2}{\Xi_i}\,.
\label{kadsarea}
\ee
The Euclidean action was also calculated in \cite{gibperpop}, and found to
be given by
\be
I = \fft{\beta {\cal A}_{n-2}}{8\pi \prod_i\Xi_i}\, 
  \Big( m - r_+^\ep\, l^{-2}\, \prod_j (r_+^2 + a_j^2)\Big)\,,
\ee
where $\beta$ is the inverse of the Hawking temperature, which is given by
\bea
\hbox{$n=$ odd}:&& 2\pi T = r_+\, (1 +r_+^2\, l^{-2})\,
  \sum_i\fft1{r_+^2 + a_i^2} -\fft1{r_+}\,,\\
\hbox{$n=$ even}:&& 2\pi T = r_+\, (1 + r_+^2\, l^{-2})\,
  \sum_i\fft1{r_+^2 + a_i^2} -\fft{1- r_+^2\, l^{-2}}{2r_+}\,.
\eea

    The traditional asymptotically-rotating Boyer-Lindquist 
coordinate system, where the angular velocities at infinity are given by
(\ref{ominf}), is related to the coordinates in (\ref{bl}) by defining
\be
\varphi_i' = \varphi_i - a_i l^{-2} t\,,\qquad t'=t\,.
\ee
The energy $E'$ calculated in the asymptotically-rotating frame, i.e. 
using the timelike Killing vector $\del/\del t'$, is
given by \cite{gibperpop}
\be
E' = E - \fft1{l^2} \sum_i a_i J_i = \fft{(n-2)m {\cal A}_{n-2}}{8\pi
              (\prod_j \Xi_j)}\,.
\ee

\end{document}